\documentclass[aip,jcp,reprint,noshowkeys,superscriptaddress]{revtex4-1}
\usepackage{graphicx,dcolumn,bm,xcolor,microtype,multirow,amscd,amsmath,amssymb,amsfonts,physics,longtable,wrapfig,bbold,xspace,siunitx}
\usepackage[version=4]{mhchem}

\usepackage[utf8]{inputenc}
\usepackage[T1]{fontenc}

\usepackage{hyperref}
\hypersetup{
    colorlinks,
    linkcolor={red!50!black},
    citecolor={red!70!black},
    urlcolor={red!80!black}
}
\usepackage{listings}
\definecolor{codegreen}{rgb}{0.58,0.4,0.2}
\definecolor{codegray}{rgb}{0.5,0.5,0.5}
\definecolor{codepurple}{rgb}{0.25,0.35,0.55}
\definecolor{codeblue}{rgb}{0.30,0.60,0.8}
\definecolor{backcolour}{rgb}{0.98,0.98,0.98}
\definecolor{mygray}{rgb}{0.5,0.5,0.5}

\definecolor{sqred}{rgb}{0.85,0.1,0.1}
\definecolor{sqgreen}{rgb}{0.25,0.65,0.15}
\definecolor{sqorange}{rgb}{0.90,0.50,0.15}
\definecolor{sqblue}{rgb}{0.10,0.3,0.60}

\lstdefinestyle{mystyle}{
    backgroundcolor=\color{backcolour},
    commentstyle=\color{codegreen},
    keywordstyle=\color{codeblue},
    numberstyle=\tiny\color{codegray},
    stringstyle=\color{codepurple},
    basicstyle=\ttfamily\footnotesize,
    breakatwhitespace=false,
    breaklines=true,
    captionpos=b,
    keepspaces=true,
    numbers=left,
    numbersep=5pt,
    numberstyle=\ttfamily\tiny\color{mygray},
    showspaces=false,
    showstringspaces=false,
    showtabs=false,
    tabsize=2
  }

  \newcolumntype{d}{D{.}{.}{-1}}

\newcommand{\mc}{\multicolumn}
\newcommand{\fnm}{\footnotemark}

\newcommand{\SupInf}{\textcolor{blue}{supporting information}}

\newcommand{\sig}{\sigma}

\newcommand{\unsafe}{1}
\newcommand{\ScHa}{2}
\newcommand{\ScHb}{3}
\newcommand{\ScOa}{4}
\newcommand{\ScOb}{5}
\newcommand{\ScOc}{6}
\newcommand{\ScFa}{7}
\newcommand{\ScFb}{8}
\newcommand{\ScFc}{9}
\newcommand{\ScFd}{10}
\newcommand{\ScSa}{11}
\newcommand{\ScSb}{12}
\newcommand{\TiNa}{13}
\newcommand{\TiNb}{14}
\newcommand{\CuHa}{15}
\newcommand{\CuHb}{16}
\newcommand{\CuHc}{17}
\newcommand{\CuFa}{18}
\newcommand{\CuClb}{19}
\newcommand{\CuFc}{20}
\newcommand{\CuCla}{21}
\newcommand{\ZnHa}{22}
\newcommand{\ZnHb}{23}
\newcommand{\ZnOa}{24}
\newcommand{\ZnOb}{25}
\newcommand{\ZnOc}{26}
\newcommand{\ZnSa}{27}

\newcommand{\LCPQ}{Laboratoire de Chimie et Physique Quantiques (UMR 5626), Universit\'e de Toulouse, CNRS, UPS, France}
\newcommand{\CEISAM}{Nantes Universit\'e, CNRS,  CEISAM UMR 6230, F-44000 Nantes, France}
\newcommand{\IUF}{Institut Universitaire de France (IUF), F-75005 Paris, France}

\begin{document}

\title{Reference Vertical Excitation Energies for Transition Metal Compounds}

\author{Denis \surname{Jacquemin}}
  \email{denis.jacquemin@univ-nantes.fr}
  \affiliation{\CEISAM}
  \affiliation{\IUF}
\author{F\'abris \surname{Kossoski}}
  \affiliation{\LCPQ}
\author{Franck \surname{Gam}}
  \affiliation{\CEISAM}
\author{Martial \surname{Boggio-Pasqua}}
  \email{martial.boggio@irsamc.ups-tlse.fr}
  \affiliation{\LCPQ}
\author{Pierre-Fran\c{c}ois \surname{Loos}}
  \email{loos@irsamc.ups-tlse.fr}
  \affiliation{\LCPQ}
\begin{abstract}
To enrich and enhance the diversity of the \textsc{quest} database of highly-accurate excitation energies [\href{https://doi.org/10.1002/wcms.1517}{V\'eril \textit{et al.}, \textit{WIREs Comput.~Mol.~Sci.}~\textbf{11}, e1517 (2021)}], we report vertical transition energies in transition metal compounds. Eleven diatomic molecules with singlet or doublet ground state containing a fourth-row transition metal (\ce{CuCl}, \ce{CuF}, \ce{CuH}, \ce{ScF}, \ce{ScH}, \ce{ScO}, \ce{ScS}, \ce{TiN}, \ce{ZnH}, \ce{ZnO}, and \ce{ZnS}) are considered and the corresponding excitation energies are computed using high-level coupled-cluster (CC) methods, namely CC3, CCSDT, CC4, and CCSDTQ, as well as multiconfigurational methods such as CASPT2 and NEVPT2. In some cases, to provide more comprehensive benchmark data, we also provide full configuration interaction estimates computed with the \textit{``Configuration Interaction using a Perturbative Selection made Iteratively''} (CIPSI) method. Based on these calculations, theoretical best estimates of the transition energies are established in both the aug-cc-pVDZ and aug-cc-pVTZ basis sets. This allows us to accurately assess the performance of CC and multiconfigurational methods for this specific set of challenging transitions. Furthermore, comparisons with experimental data and previous theoretical results are also reported.
\bigskip
\begin{center}
  \boxed{\includegraphics[width=0.5\linewidth]{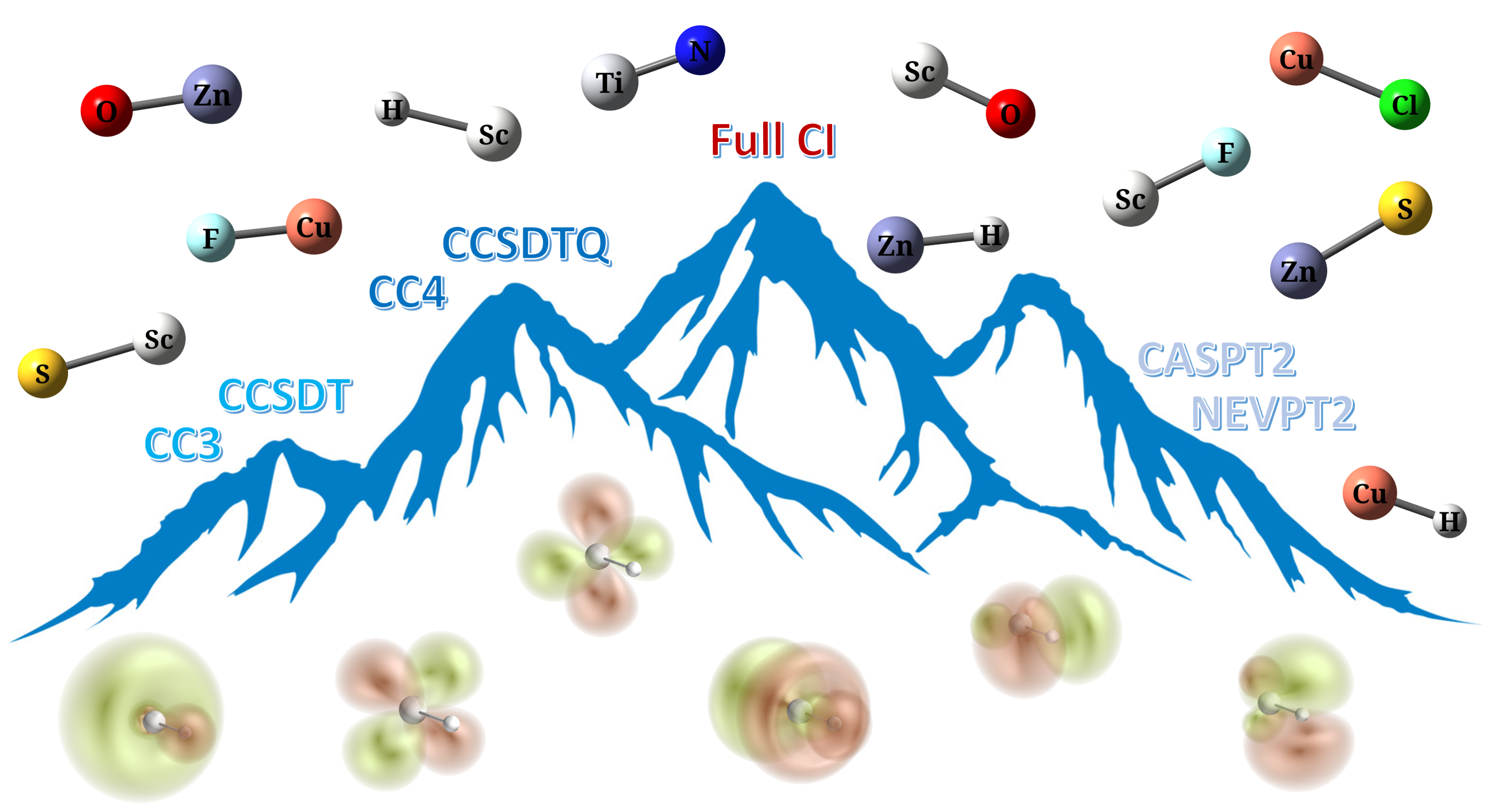}}
\end{center}
\bigskip
\end{abstract}

\maketitle

\section{Introduction}
\label{sec:introduction}
Understanding the electronic structure of transition metal compounds \cite{Khomskii_2014} is critical for unraveling their specific behaviors and applications in a wide range of fields, such as chemistry and biology. \cite{Trautwein_1997} Their electronic structure is characterized by the presence of partially filled $d$ orbitals in the transition metal atoms, which gives rise to their unique properties such as variable oxidation states, \cite{Askerka_2016} magnetic behavior, \cite{Ebert_1996,vanOudenaarden_1998,Gerhard_2010} and catalytic activity. \cite{Melius_1976} The empty $d$ orbitals can participate in chemical reactions, \cite{Bolm} allowing for the transfer of electrons during redox processes. \cite{Ma_2021} Transition metal catalysts find applications in various industrial processes, including hydrogenation, oxidation, and carbon-carbon bond formation. \cite{GarcaMelchor_2013,Guest_2015,Sperger_2015}

From a general perspective, investigating molecular excited states is essential for understanding their reactivity, photophysical properties, and catalytic behavior. Indeed, the presence of the excited electron modifies the electronic structure of the molecule, affecting reactions such as bond activation, insertion, or reductive processes. \cite{Bernardi_1990,Klessinger_1995,Robb_2007,Delgado_2010,Olivucci_2010} The excited states of molecules containing transition metals have very peculiar characteristics and reactivity due to the presence of the transition metal atom.
For example, in photocatalysis, the absorption of light can lead to the formation of reactive excited states that participate in photochemical reactions. \cite{Twilton_2017}

Experimentally characterizing excited states in transition metal compounds is challenging due to their often short lifetimes and low transition probabilities. Transient absorption spectroscopy, \cite{Geneaux_2019} time-resolved techniques, \cite{Haglund_2007} and advanced spectroscopic methods are required to observe and analyze the excited-state behavior. Additionally, the identification and assignment of the observed spectral features are challenging due to the complexity of the excited-state manifold in these systems. \cite{Bokarev_2019}

From a theoretical point of view, the study of excited states in transition metal compounds presents several challenges due to their complex electronic structures and intricate interactions. \cite{Harrison_2000,Harvey_2006,Zhao_2006,Sperger_2016,Santoro_2016b,Friesner_2017,Bokarev_2019,Hait_2019} First, they often require sophisticated theoretical methods to accurately describe the electronic structure of both their ground state and their excited states. \cite{Roos_1996a,Piecuch_2002,Dreuw_2005,Krylov_2006,Sneskov_2012,Gonzales_2012,Laurent_2013,Adamo_2013,Ghosh_2018,Blase_2020,Loos_2020a} These calculations are computationally demanding and require the use of high-level quantum chemical approaches, \cite{Helgakerbook} such as multireference methods, \cite{Khedkar_2021} to account for the strong correlation effects present in many transition metal systems (see below). \cite{Buendia_2013,Aoto_2017,Shee_2019,Wagner_2003,Wagner_2007,Diedrich_2005,Caffarel_2005,Caffarel_2014,Doblhoff_Dier_2016,Scemama_2018a,Giner_2018a,Suo_2017,Nooijen_2000,Korbel_2014} Second, the large number of electrons and basis functions involved in these calculations further add to the computational complexity. Third, transition metal compounds often exhibit multiple spin states, which affects not only their reactivity but also their magnetic properties. \cite{Wang_2011} This leads to the accumulation of excited states with potentially the same spin and spatial symmetries in a very narrow energy window, which complicates the interpretation of experimental results and theoretical calculations. \cite{Loos_2020a,Veril_2021}

As mentioned above, transition metal derivatives often exhibit strong (or static) correlation, which refers to the intricate interactions among electrons occupying the $d$ orbitals. Strong correlation is often a signature of the multiconfigurational character of the electronic wave function, meaning that the ground state and/or excited states cannot be accurately described by a single Slater determinant (single-reference wave function) but require a linear combination of multiple determinants (multireference wave function) to accurately capture the electronic structure. The remaining dynamic correlation must also be taken into account as it systematically plays a crucial role in describing the excited states.

To accurately describe the multiconfigurational character and strong electron correlation, methods based on configuration interaction (CI) are commonly employed. \cite{SzaboBook} This class of methods allows for the mixing of different electronic configurations and provides a flexible framework to capture the electronic correlation effects. If one considers all possible electronic configurations, the resulting full CI (FCI) wave function corresponds to the exact solution of the Schr\"odinger equation within a given one-electron basis set. Unfortunately, the ensemble of these configurations, known as the Hilbert space, has a size that grows exponentially fast with the system size, leading to a prohibitive computational cost in most applications.

Multiconfigurational self-consistent field methods, such as complete-active-space self-consistent field (CASSCF), account for all determinants generated by distributing a given number of electrons in a given number of active orbitals, therefore incorporating, by design, static correlation. Besides, unlike in CI, orbitals are variationally optimized. The missing dynamic correlation is usually recovered via low-order perturbation theory as in the complete-active-space second-order perturbation theory (CASPT2) \cite{Andersson_1990,Andersson_1992,Roos_1995a} or the $N$-electron valence state second-order perturbation theory (NEVPT2). \cite{Angeli_2001a,Angeli_2001b,Angeli_2002,Angeli_2006}
For CASSCF-based methods, selecting an appropriate active space is critical for capturing the important electron correlation effects while keeping the computational cost manageable. In transition metal compounds, the active space typically involves the $d$ orbitals of the metal center and the orbitals involved in the ligand interactions. Choosing an appropriate active space is a delicate balance between including a sufficient number of active orbitals to describe the correlation effects and keeping a reasonable computational cost.

Coupled-cluster (CC) methods offer an alternative approach, based on an exponential ansatz of the wave function, that allows for size-extensive and systematically improvable calculations towards the FCI limit. These methods exhibit polynomial scaling and have been extensively studied in the literature. \cite{Cizek_1966,Cizek_1969,Paldus_1992,Crawford_2000,Bartlett_2007,Shavitt_2009} Coupled-cluster methods systematically incorporate higher levels of excitations to improve accuracy. For instance, CC with singles and doubles (CCSD), \cite{Purvis_1982,Scuseria_1987,Koch_1990a,Stanton_1993a,Stanton_1993b} CC with singles, doubles, and triples (CCSDT), \cite{Noga_1987,Scuseria_1988,Watts_1994,Kucharski_2001} and CC with singles, doubles, triples, and quadruples (CCSDTQ) \cite{Kucharski_1991,Kallay_2001,Hirata_2004,Kallay_2003,Kallay_2004a} can be obtained by successively adding higher excitation levels. The computational cost of these methods scales as $\order*{N^6}$, $\order*{N^8}$, and $\order*{N^{10}}$, respectively. Moreover, to reduce computational expenses, each of these methods can be approximated by the CC$n$ family of methods. This family includes CC2 ($\order*{N^5}$), \cite{Christiansen_1995a,Hattig_2000} CC3 ($\order*{N^7}$), \cite{Christiansen_1995b,Koch_1995,Koch_1997,Hald_2001,Paul_2021} and CC4 ($\order*{N^9}$). \cite{Kallay_2004b,Kallay_2005,Matthews_2015,Loos_2021a,Loos_2022a} These variants provide cost-effective alternatives while still maintaining acceptable levels of accuracy compared to their ``complete'' variant. Excited-state energies and properties can be obtained within the CI framework by searching for higher roots of the CI matrix and their corresponding eigenvectors. Similarly, at CC level, one can access excited states using the equation-of-motion (EOM) \cite{Rowe_1968a,Emrich_1981,Sekino_1984,Geertsen_1989,Stanton_1993a,Comeau_1993,Watts_1994} or linear-response (LR) \cite{Monkhorst_1977,Dalgaard_1983,Sekino_1984,Koch_1990c,Koch_1990a} formalisms.

\section{The \textsc{quest} database}
\label{sec:quest}

Benchmark sets and their corresponding reference data serve as a cornerstone in electronic structure theory, supporting the development, validation, and improvement of computational methods for both the ground state \cite{Pople_1989,Curtiss_1991,Curtiss_1997,Curtiss_1998,Curtiss_2007,Jureka_2006,Rezac_2011,vanSetten_2015,Stuke_2020,Tajti_2004,Bomble_2006,Harding_2008,Motta_2017,Williams_2020,Zhao_2006,Mardirossian_2017,Goerigk_2010,Goerigk_2011a,Goerigk_2011b,Goerigk_2017} and the excited states. \cite{Schreiber_2008,Silva-Junior_2008,Silva-Junior_2010,Silva-Junior_2010b,Silva-Junior_2010c,Leang_2012,Schwabe_2017,Casanova-Paez_2019,Casanova_Paes_2020,Furche_2002,Send_2011a,Winter_2013,Dierksen_2004,Goerigk_2010a,Jacquemin_2012,Jacquemin_2015b,Kozma_2020} In the context of molecular excited states, a benchmark set refers to a collection of molecules with known reference data that is used to evaluate the accuracy, reliability, and limitations of computational methods in predicting the properties of electronic excited states, such as excitation energies, oscillator strengths, transition dipole moments, and other spectroscopic observables.

They provide a standardized framework for evaluating the performance of different computational methods in predicting excited-state properties, contributing to the reproducibility and transparency of computational studies in the field. By comparing different methods against each other, researchers can identify the strengths and weaknesses of different approaches and gain insights into their limitations, hence guiding the development of new computational methods for electronic excited states. Besides, these investigations also assist researchers and practitioners in selecting appropriate methods for specific applications or molecules.

The reference data are typically obtained from highly-accurate theoretical calculations or experimental measurements. Their accuracy and reliability are mandatory for ensuring the meaningfulness of the benchmarking process. In some cases, reference data may be obtained from experiments, such as spectroscopic measurements or photochemical data, but these experimental values are not always available or may have uncertainties. \cite{Loos_2019b} The selection of molecules for a benchmark set aims to cover a diverse range of electronic structures and properties, including different types of excited states, as well as a variety of chemical environments and molecular sizes. Importantly, it should also include challenging cases that test the capabilities of the methods under investigation.

\begin{figure}
  \includegraphics[width=\linewidth]{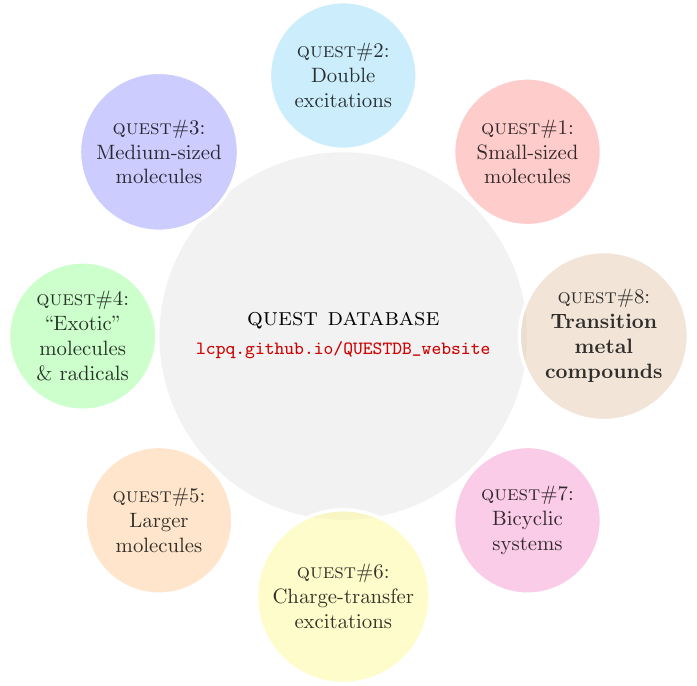}
  \caption{The 8 subsets composing the \textsc{quest} database of highly-accurate excitation energies with the inclusion of a new subset gathering excited states of transition metal compounds (\textsc{quest}\#8).}
  \label{fig:QUEST}
\end{figure}

Since 2018, our research groups have made significant efforts to develop a comprehensive and diverse database of highly-accurate vertical excitation energies for small- and medium-sized (organic) molecules. This database, named \textsc{quest}, \cite{Loos_2020d,Veril_2021} has been meticulously curated and expanded over time. It currently comprises seven subsets, as illustrated in Fig.~\ref{fig:QUEST}:
\begin{itemize}
  \item \textsc{quest}\#1: This subset consists of 110 vertical excitation energies (and oscillator strengths) in small molecules containing 1 to 3 non-hydrogen atoms. \cite{Loos_2018a} Primarily focused on singly-excited states, the theoretical best estimates (TBEs) were determined using FCI calculations.
  \item \textsc{quest}\#2: Comprising 20 vertical transition energies for doubly-excited states in 14 small and medium-sized molecules, \cite{Loos_2019c} this subset relied predominantly on FCI calculations to define the TBEs, except for the largest molecules in the set.
  \item \textsc{quest}\#3: This subset encompasses 238 excitation energies (and oscillator strengths) for 27 medium-sized molecules containing 4 to 6 non-hydrogen atoms. \cite{Loos_2020c} The TBEs in this subset were originally defined using CCSDT or CCSDTQ methods, and more recent improvements have been made with CC4 and CCSDTQ approaches. \cite{Loos_2021c,Loos_2022b}
  \item \textsc{quest}\#4: Composed of two distinct parts, this subset includes an ``exotic'' subset of 30 vertical excitation energies for closed-shell molecules containing F, Cl, P, and Si atoms, and a ``radical'' subset of 51 doublet-doublet transitions in 24 small open-shell molecules. \cite{Loos_2020f} In total, there are 81 TBEs, mostly obtained with FCI.
  \item \textsc{quest}\#5: Featuring 80 excitation energies in 13 (mostly large) molecules, this subset mostly contains TBE calculations at the CCSDT level. \cite{Veril_2021}
  \item \textsc{quest}\#6: Specifically designed for the study of intramolecular charge-transfer transitions, this subset provides highly-accurate vertical excitation energies for 30 such transitions in 17 $\pi$-conjugated compounds obtained at the CCSDT level. \cite{Loos_2021a}
  \item \textsc{quest}\#7: This subset contains 91 vertical excitation energies for 10 bicyclic molecules computed at the CC3 or CCSDT levels. \cite{Loos_2021b}
\end{itemize}

As evidenced by the above description, \textsc{quest} employs FCI and high-order CC methods to generate highly-accurate reference data in triple-$\zeta$ basis, alongside additional basis set corrections when possible. In most cases, geometry optimization has been carried out at the CC3/aug-cc-pVTZ level. A significant advantage of the \textsc{quest} dataset is its independence from experimental values, eliminating potential biases associated with experiments and facilitating direct theoretical comparisons. The employed protocol ensures uniformity, enabling straightforward cross-comparisons. This approach allowed the benchmarking of a wide range of excited-state wave function methods, including those accounting for double and triple excitations, as well as multiconfigurational methods. Furthermore, chemically-accurate theoretical 0-0 energies have been computed, providing a more direct comparison to experimental data. \cite{Loos_2018c,Loos_2019a,Loos_2019b} Presently, our ongoing efforts are dedicated to obtaining highly-accurate excited-state properties such as dipoles and oscillator strengths for small and medium-sized molecules. \cite{Chrayteh_2021,Sarkar_2021,Sarkar_2022,Damour_2022}

The creation of the \textsc{quest} dataset involved a very significant computational effort, with numerous calculations performed for each of the associated articles. \cite{Loos_2018a,Loos_2019c,Loos_2020c,Loos_2020f,Veril_2021,Loos_2021a,Loos_2021b} To access and manipulate the data, a web application has been developed and hosted on a GitHub repository (\url{https://github.com/LCPQ/QUESTDB_website}). The web application can be accessed at \url{https://lcpq.github.io/QUESTDB_website}, providing users with the ability to plot statistical indicators for selected subsets of molecules, methods, and basis sets.

The utilization of the \textsc{quest} database as a benchmark for excited-state methods has gained attraction among research groups worldwide.
For instance, the database has been employed to assess orbital-optimized DFT for double excitations, \cite{Hait_2020,Hait_2021} multistate DFT, \cite{Zhu_2023} and TD-DFT. \cite{Liang_2022} Additionally, it has facilitated the evaluation of hybrid \cite{Grotjahn_2021} and double hybrid \cite{Casanova_2019,Casanova_2021,Mester_2021a,Mester_2021b} functionals, quantum Monte Carlo methods for excited states \cite{Dash_2019,Otis_2020,Dash_2021,Shepard_2022,Otis_2023}, multiconfiguration methods \cite{Sarkar_2022,Boggio-Pasqua_2022,King_2022,Wang_2022} and others. \cite{Gould_2022,Kossoski_2023,Dombrowski_2023,Kossoski_2023a} These studies demonstrate the widespread use of the \textsc{quest} database as a valuable resource for the rigorous assessment of excited-state methods.

In this study, we aim to enhance the diversity of our database and incorporate chemically challenging cases. Specifically, we perform excited-state calculations for 11 diatomic molecules with a singlet or doublet ground state, each containing a fourth-row transition metal: \ce{CuCl}, \ce{CuF}, \ce{CuH}, \ce{ScF}, \ce{ScH}, \ce{ScO}, \ce{ScS}, \ce{TiN}, \ce{ZnH}, \ce{ZnO}, and \ce{ZnS}. To establish highly-accurate reference vertical excitation energies, we determine TBEs using a combination of FCI and CCSDTQ data in the aug-cc-pVDZ and aug-cc-pVTZ basis. Leveraging these reference values, we conduct a comprehensive assessment of lower-order CC methods, namely, CC3, CCSDT, and CC4, as well as benchmark the performance of both CASPT2 and NEVPT2 for this set of excitations. This contributes to a more thorough understanding of the capabilities and limitations of these computational methods in addressing the electronic excitations of the aforementioned diatomic molecules.

Spin-orbit coupling, which arises from the relativistic effects on the electrons' motion, is important in transition metal compounds as it couples the spin states of the electrons and affects the energy ordering and mixing of the excited states. Properly accounting for relativistic effects is crucial when performing experiment \textit{vs} theory comparisons, and it requires specified theoretical approaches to accurately describe the electronic structure and energetics. Here, because our aim is to rely solely on theoretical values and to perform theory \textit{vs} theory comparisons, we eschew taking into account relativistic effects. We refer the interested reader to Ref.~\onlinecite{Aoto_2017} for a state-of-the-art treatment of these systems based on single- and multi-reference CC methodologies that incorporate core-valence and relativistic effects as well as complete basis set extrapolations.

\section{Computational Methodology}
\label{sec:comp_det}
The ground-state geometries of the singlet and doublet states have been optimized, in the frozen-core approximation at the CC3/aug-cc-pVTZ and UCCSD(T)/aug-cc-pVTZ levels of theory, respectively. These calculations were performed with \textsc{cfour} \cite{Matthews_2020} and \textsc{gaussian16}, \cite{g16} respectively. Large frozen cores have been systematically selected. (Additional calculations for small cores can be found in the {\SupInf} showing that the deviations between small and large frozen-core excitation energies are small.) The optimized bond lengths are reported in Table \ref{tab:geom} alongside the electronic ground-state symmetry of each system and experimental values extracted from Ref.~\onlinecite{Aoto_2017}. For all the systems considered, we performed calculations using two diffuse-containing Gaussian basis sets (aug-cc-pVDZ and aug-cc-pVTZ).

FCI vertical excitation energies were obtained with selected CI calculations \cite{Giner_2013,Giner_2015,Liu_2016a,Holmes_2017,Mussard_2018,Tubman_2018,Chien_2018,Tubman_2020,Loos_2020i,Yao_2020,Zhang_2020,Damour_2021,Yao_2021,Zhang_2021,Larsson_2022,Coe_2022} based on the \textit{Configuration Interaction using a Perturbative Selection made Iteratively} (CIPSI) algorithm. \cite{Huron_1973} All these calculations were performed with \textsc{quantum package} \cite{Garniron_2019} following the same protocol as our previous studies. \cite{Loos_2018a,Loos_2019c,Loos_2020c} Extrapolation errors are estimated following the procedure of Ref.~\onlinecite{Veril_2021}.

For the singlet excited states of closed-shell systems, the CC calculations were carried out using \textsc{cfour}, \cite{Matthews_2020} which offers an efficient implementation of high-order CC methods up to quadruples. \cite{Matthews_2015} For the triplet excited states of closed-shell derivatives, we relied on \textsc{psi4} \cite{psi4-jcp} for the (U-)CC3 calculations and \textsc{mrcc} \cite{Kallay_2020} for the (U-)CCSDT and (U-)CCSDTQ calculations. For the open-shell transition metal derivatives, the latter two codes were used as well for the corresponding CC calculations, that were achieved starting from the restricted open-shell Hartree-Fock (ROHF) solution.

The multiconfigurational calculations were performed using a state-averaged (SA) CASSCF wave function, which included the ground state and, at least, the excited states of interest. Additional excited states were included in some cases to address convergence and root-flipping issues. The CASPT2 calculations were performed within the RS2 contraction scheme (unless otherwise stated), as implemented in \textsc{molpro}, \cite{Werner_2020} with a default IPEA shift of \SI{0.25}{\hartree}.\cite{Ghigo_2004,Zobel_2017} To mitigate the intruder state problem, a level shift of \SI{0.3}{\hartree} was systematically applied. \cite{Roos_1995b,Roos_1996b} In some cases, we have also performed partially-contracted (PC) NEVPT2 calculations, as well as CASPT2 calculations without IPEA shift (labeled as NOIPEA). These additional data can be found in the {\SupInf} where one would also find strongly-contracted (SC) NEVPT2 results. For each system and transition, the {\SupInf} also provides a detailed description of the active spaces for each symmetry representation.

\begin{table}
\caption{Electronic ground-state symmetry and corresponding bond length (in \si{\angstrom}) of the 11 diatomic molecules considered herein.}
\label{tab:geom}
\begin{ruledtabular}
\begin{tabular}{lccc}
          &  Electronic    &  \mc{2}{c}{Bond Length (\si{\angstrom})} \\
          						\cline{3-4}
  System  &  Ground State  &  This Work  & Exp. (Ref.~\onlinecite{Aoto_2017}) \\
  \hline
  ScH    &  $1\,^1\Sigma^+$  & 1.796 & 1.7754 \\
  ScO    &  $1\,^2\Sigma^+$  & 1.699 & 1.6661 \\
  ScF    &  $1\,^1\Sigma^+$  & 1.788 & 1.787  \\
  ScS    &  $1\,^2\Sigma^+$  & 2.168 & 2.1353 \\
  TiN    &  $1\,^2\Sigma^+$  & 1.599 & 1.5802 \\
  CuH    &  $1\,^1\Sigma^+$  & 1.480 & 1.4626 \\
  CuF    &  $1\,^1\Sigma^+$  & 1.758 & 1.7449 \\
  CuCl   &  $1\,^1\Sigma^+$  & 2.075 & 2.0512 \\
  ZnH    &  $1\,^2\Sigma^+$  & 1.603 & 1.5935 \\
  ZnO    &  $1\,^1\Sigma^+$  & 1.700 & 1.7047 \\
  ZnS    &  $1\,^1\Sigma^+$  & 2.068 & 2.0464 \\
\end{tabular}
\end{ruledtabular}
\end{table}

\section{Results and Discussion}
\label{sec:discussion}

\begin{figure*}
  \includegraphics[width=\linewidth]{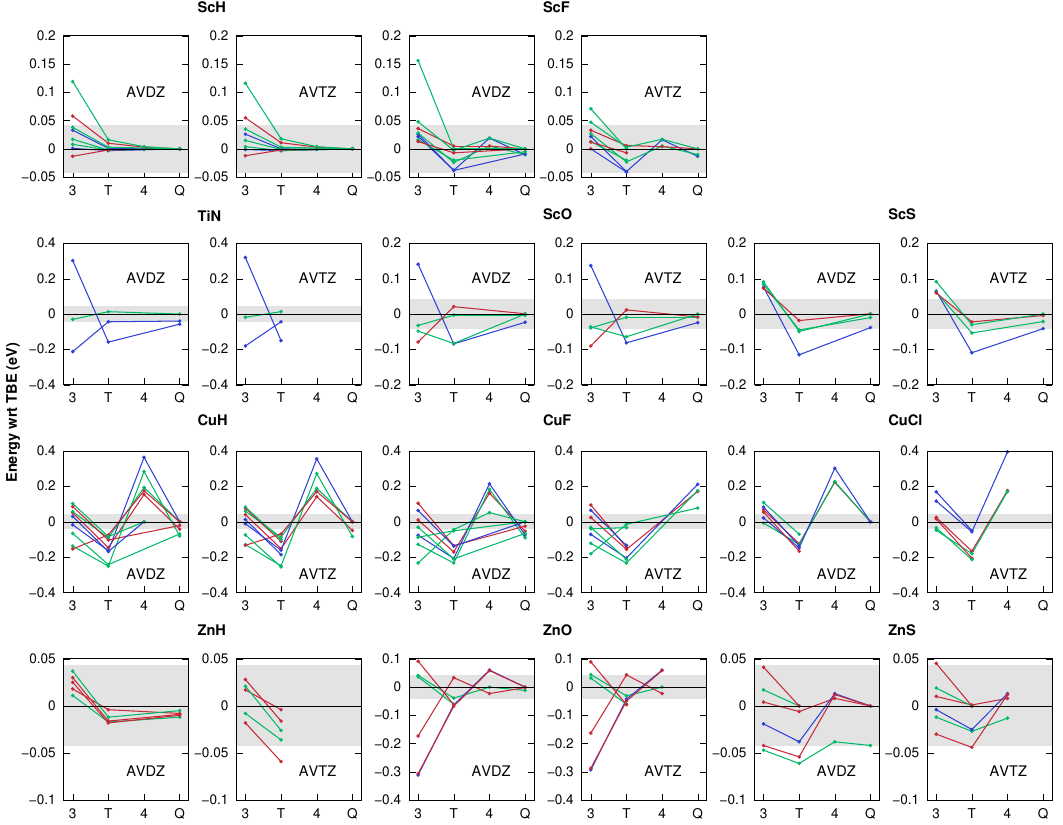}
    \caption{Convergence of the errors in excitation energies (with respect to the TBEs) along the coupled-cluster models (CC3, CCSDT, CC4, and CCSDTQ, denoted by their last digit in the figure),
    with the aug-cc-pVDZ (AVDZ) and aug-cc-pVTZ (AVTZ) basis set, for $\Sigma$ (red), $\Pi$ (green), and $\Delta$ (blue) excited states of transition metal diatomics.
    The gray shaded region indicates deviations of \SI{\pm0.043}{\eV} with respect to the TBE.}
  \label{fig:conv}
\end{figure*}

The results for the singlet and triplet transitions are reported in Table \ref{tab:singlets}, whereas those for the doublet transitions are reported in Table \ref{tab:doublets}. Table \ref{tab:TBE} contains our TBEs, along with selected available results from the literature. To the best of our knowledge, these are usually the most up-to-date theoretical or experimental data for each state. The interested reader can consult the corresponding references to find more exhaustive comparisons with prior results, which is not our focus here. The convergence of the CC energies towards the TBEs is shown in Fig.~\ref{fig:conv}. We consider 22 out of our 67 TBEs to be unsafe (meaning errors potentially greater than \SI{1}{kcal/mol} or \SI{0.043}{\eV}). Despite the uncertainties for this subset of TBEs, one can still gauge the convergence profile of the CC series, since a new TBE would only set a new reference energy.

It is important to bear in mind that, in most cases, our computed vertical excitation energies are not directly comparable to the previously reported data shown in Table \ref{tab:TBE}. There are three reasons for that. First, ground-state geometries might be slightly different. Second, there are differences between the Hamiltonian employed in the calculations and the true physical Hamiltonian. Here, we adopt the non-relativistic Coulombic Hamiltonian, which neglects spin-orbit coupling and relativistic effects. These effects are important for transition metal compounds, and should be taken into account had the goal been to obtain a quantitative comparison with experimental observables. Ignoring them is justifiable though, because we are interested in obtaining accurate non-relativistic excitation energies, which are far greater in magnitude than the contribution from the above-mentioned effects. For this reason, when comparing the present results with previous calculations, we present those that similarly ignore relativistic effects, when available.

A third aspect is that the excitation energies listed in Table \ref{tab:TBE} often correspond to different observables. Experiments typically report specific vibronic transitions, particularly for those between vibrational ground states of electronic ground and excited states, the so-called 0-0 energies, also referred to as $T_0$. In turn, theoretical studies usually present potential energy curves, from which the minimum energy separation between ground and excited states (the adiabatic or $T_e$ energy) is obtained. Modeling the vibrational levels would provide information about the 0-0 energy, which can be compared with experimental values. \cite{Dierksen_2004,Jacquemin_2012,Winter_2013,Jacquemin_2015b,Loos_2018c,Loos_2019a,Loos_2019b} Here, instead, we provide very accurate vertical excitation energies. Considering non-relativistic potential energy curves, our vertical value represents an upper bound for $T_e$ and would also be expected to be higher than the 0-0 value in most cases.

In the following, we discuss in detail each transition metal compound, grouped into different families. When comparing the available data with our TBEs, we always refer to the aug-cc-pVTZ values. Next, we present the global view of our full set of results, discussing the performance of the different methodologies for transition metal diatomics and comparing them with previous subsets of the \textsc{quest} database devoted to organic compounds. \cite{Veril_2021}

\begin{squeezetable}
\begin{table*}
\caption{Vertical excitation energies (in \si{\eV}) of the lowest singlet and triplet excited states of \ce{ScH}, \ce{ScF}, \ce{CuH}, \ce{CuF}, \ce{CuCl}, \ce{ZnO}, and \ce{ZnS} at various levels of theory. 3, T, 4, Q, CAS, and NEV stand for CC3, CCSDT, CC4, CCSDTQ, CASPT2(IPEA), and PC-NEVPT2, respectively. The CASPT2 values for \ce{CuF} are computed in the R2C formalism.}
\label{tab:singlets}
\begin{ruledtabular}
\begin{tabular}{llcccccccccccccc}
  &          &  \mc{7}{c}{aug-cc-pVDZ}          &  \mc{7}{c}{aug-cc-pVTZ} \\
                     \cline{3-9}                        \cline{10-16}
  Mol.&  State          &  3    &  T  &  4    &  Q  &  CAS  &  NEV    &  FCI      &  3    &  T  &  4    & Q  &  CAS  &  NEV  &  FCI \\
  \hline
  ScH &  $1\,^1\Delta$(4s,3d)        &  0.639  &  0.607  &  0.606  &  0.606  &  0.519  &  0.538  &  0.606(1)  &  0.698  &  0.672  &  0.671  &  0.672  &  0.515  &  0.535  &  0.671(1) \\
      &  $1\,^1\Pi$(4s,3d)           &  0.837  &  0.818  &  0.819  &  0.820  &  0.783  &  0.786  &  0.820(1)  &  0.890  &  0.874  &  0.875  &  0.875  &  0.764  &  0.769  &  0.875(1) \\
      &  $2\,^1\Sigma^+$(4s,3d)      &  1.894  &  1.846  &  1.839  &  1.836  &  2.060  &  2.003  &  1.836(1)  &  1.900  &  1.856  &  1.848  &  1.845  &  2.040  &  1.983  &           \\
      &  $2\,^1\Pi$(4s$^2$,3d$^2$)           &  2.300  &  2.197  &  2.185  &  2.181  &  2.122  &  2.141  &  2.181(1)  &  2.331  &  2.233  &  2.219  &  2.215  &  2.129  &  2.149  &           \\
      &  $1\,^3\Delta$(4s,3d)        &  0.365  &  0.361  &         &  0.364  &  0.278  &  0.292  &  0.363(1)  &  0.445  &  0.443  &         &  0.446  &  0.280  &  0.295  &  0.445(1) \\
      &  $1\,^3\Pi$(4s,3d)           &  0.573  &  0.563  &         &  0.565  &  0.520  &  0.532  &  0.565(1)  &  0.625  &  0.618  &         &  0.621  &  0.510  &  0.523  &  0.620(1) \\
      &  $1\,^3\Sigma^+$(4s,3d)      &  0.807  &  0.818  &         &  0.820  &  0.821  &  0.834  &  0.820(1)  &  0.840  &  0.850  &         &  0.852  &  0.813  &  0.827  &  0.851(1) \\
      &  $2\,^3\Pi${(4s,3d)}           &  1.870  &  1.835  &         &  1.832  &  1.845  &  1.832  &            &  1.852  &  1.820  &         &  1.817  &  1.853  &  1.839  &           \\
  \hline
  ScF &  $1\,^1\Delta$(4s,3d)      &  0.823  &  0.759  &  0.816  &  0.787  &  0.559  &  0.612  &  0.797(1)  &  0.925  &  0.863  &  0.919  &  0.890  &  0.554  &  0.584  &  0.903(4)  \\
      &  $1\,^1\Pi$(4s,3d)         &  1.602  &  1.550  &  1.579  &  1.564  &  1.353  &  1.406  &  1.574(3)  &  1.698  &  1.648  &  1.676  &  1.661  &  1.344  &  1.409  &  1.653(26) \\
      &  $2\,^1\Sigma^+$(4s,3d)    &  2.382  &  2.351  &  2.351  &  2.346  &  2.378  &  2.385  &  2.356(18) &  2.437  &  2.410  &  2.408  &  2.404  &  2.366  &  2.382  &            \\
      &  $2\,^1\Pi$(4s$^2$,3d$^2$) &  2.806  &  2.756  &  2.777  &  2.758  &  2.479  &  2.562  &            &  2.884  &  2.840  &  2.854  &  2.837  &  2.478  &  2.569  &            \\
      &  $1\,^3\Delta$(4s,3d)      &  0.550  &  0.490  &         &  0.519  &  0.292  &  0.341  &  0.528(2)  &  0.665  &  0.606  &         &         &  0.292  &  0.318  &  0.647(3)  \\
      &  $1\,^3\Pi$(4s,3d)         &  1.053  &  1.017  &         &  1.032  &  0.896  &  0.923  &  1.037(4)  &  1.122  &  1.089  &         &         &  0.896  &  0.933  &            \\
      &  $1\,^3\Sigma^+$(4s,3d)    &  1.414  &  1.394  &         &  1.401  &  1.389  &  1.403  &            &  1.471  &  1.452  &         &         &  1.389  &  1.408  &            \\
      &  $2\,^3\Pi$(4s$^2$,3d$^2$) &  2.625  &  2.469  &         &  2.469  &  2.409  &  2.478  &            &  2.664  &  2.593  &         &         &  2.422  &  2.503  &            \\
  \hline
  CuH &  $2\,^1\Sigma^+$(3d,4s)            &  3.101  &  2.905  &    3.205  &    3.009  &  3.102  &  &  3.051(4)  &  3.120  &    2.928  &    3.222  &    3.031  &    3.143  &  &  3.080(7)  \\
      &  $1\,^1\Delta$(3d,4s) &  3.701  &  3.554  &    4.082  &    3.718  &  3.558  &  &            &  3.743  &    3.593  &    4.110  &    3.754  &    3.614  &  &            \\
      &  $1\,^1\Pi$(3d,4s)         &  3.651  &  3.537  &    4.072  &    3.707  &  3.571  &  &  3.788(26) &  3.700  &    3.580  &    4.102  &    3.746  &    3.623  &  &  3.829(28) \\
      &  $2\,^1\Pi$(3d,4s)        &  5.527  &  5.379  &    5.664  &    5.470  &  5.408  &  &            &  5.557  &    5.410  &    5.686  &    5.497  &    5.449  &  &            \\
      &  $3\,^1\Sigma^+$(3d,4p) &  5.560  &  5.639  &    5.888  &    5.713  &  5.971  &  &            &  5.618  &    5.682  &    5.922  &    5.751  &    6.038  &  &            \\
      &  $1\,^3\Sigma^+$(3d,4s)            &  2.600  &  2.411  &           &    2.493  &  2.536  &  &  2.514(5)  &  2.626  &    2.439  &           &           &    2.572  &  &  2.539(37) \\
      &  $1\,^3\Pi$(3d,4s)         &  3.456  &  3.280  &           &    3.453  &  3.333  &  &  3.522(11) &  3.508  &    3.327  &           &           &    3.393  &  &  3.561(37) \\
      &  $1\,^3\Delta$(3d,4s) &  3.591  &  3.393  &           &    3.561  &  3.422  &  &            &  3.631  &    3.432  &           &           &    3.479  &  &            \\
      &  $2\,^3\Pi$(3d,4s)        &  4.865  &  4.678  &           &    4.762  &  4.691  &  &            &  4.900  &    4.716  &           &           &    4.746  &  &            \\
  \hline
  CuF &  $2\,^1\Sigma^+$(3d,4s)            &  2.571  &    2.391  &    2.723  &    2.508    &  2.528  &  &    2.561(10) &  2.596    &  2.416  &    2.744  &  &      2.539  &  &  2.604(28) \\
      &  $1\,^1\Pi$(3d,4s)         &  2.623  &    2.518  &    2.934  &    2.660    &  2.632  &  &    2.751(8)  &  2.671    &  2.560  &    2.970  &  &      2.674  &  &            \\
      &  $1\,^1\Delta$(3d,4s) &  3.209  &    3.077  &    3.500  &    3.211    &  3.014  &  &    3.285(49) &  3.245    &  3.111  &    3.528  &  &      3.045  &  &            \\
      &  $2\,^1\Pi$(3d,4s)        &  5.681  &    5.870  &    5.966  &    5.914    &  6.028  &  &              &  5.752    &  5.919  &    6.011  &  &      6.047  &  &            \\
      &  $1\,^3\Sigma^+$(3d,4s)            &  2.121  &    1.882  &           &    1.992    &  1.941  &  &    2.017(20) &  2.158    &  1.918  &           &  &      1.987  &  &  2.066(58) \\
      &  $1\,^3\Pi$(3d,4s)         &  2.390  &    2.211  &           &    2.355    &  2.317  &  &    2.421(28) &  2.439    &  2.258  &           &  &      2.367  &  &            \\
      &  $1\,^3\Delta$(3d,4s) &  3.001  &    2.799  &           &    2.937    &  2.758  &  &              &  3.037    &  2.838  &           &  &      2.793  &  &            \\
      &  $2\,^3\Pi$(3d,4s)        &  5.679  &    5.715  &           &    5.766    &  5.791  &  &              &  5.760    &  5.767  &           &  &      5.825  &  &            \\
  \hline
 CuCl &  $2\,^1\Sigma^+$($\sig$,4s)  &  3.072  &    2.880  &    3.227  &    3.004  &  3.088  &  3.333  &  &  3.090  &    2.898  &    3.242  &  &  3.149  &  3.030  &  \\
      &  $1\,^1\Pi$(3d,4s)   &  3.001  &    2.880  &    3.235  &    3.008  &  2.965  &  2.971  &  &  3.044  &    2.912  &    3.263  &  &  3.049  &  3.015  &  \\
      &  $1\,^1\Delta$(3d,4s)    &  3.562  &    3.392  &    3.843  &    3.540  &  3.533  &  3.539  &  &  3.596  &    3.422  &    3.874  &  &  3.472  &  3.454  &  \\
      &  $1\,^3\Sigma^+$($\sig$,4s)  &  2.681  &    2.460  &           &           &  2.626  &  2.529  &  &  2.708  &    2.486  &           &  &  2.692  &  2.578  &  \\
      &  $1\,^3\Pi$(3d,4s)   &  2.770  &    2.591  &           &           &  2.661  &  2.654  &  &  2.811  &    2.631  &           &  &  2.845  &  2.818  &  \\
      &  $1\,^3\Delta$(3d,4s)    &  3.364  &    3.145  &           &           &  3.281  &  3.277  &  &  3.398  &    3.179  &           &  &  3.229  &  3.203  &  \\
  \hline
  ZnO &  $1\,^1\Pi$(2p,4s)          &  0.812  &  0.732  &  0.771  &  0.759  &  0.523  &  0.556  &  0.771(2)  &  0.835  &  0.760  &  0.791  &  &  0.530  &  0.557  & \\
      &  $2\,^1\Sigma^+$($\sig$,4s) &  3.244  &  3.450  &  3.394  &  3.415  &  3.767  &  3.720  &  3.417(22) &  3.260  &  3.466  &  3.400  &  &  3.763  &  3.718  & \\
      &  $1\,^1\Delta$(2p,4p)       &  4.300  &  4.548  &  4.671  &  4.611  &  4.365  &  4.399  &            &  4.352  &  4.602  &  4.705  &  &  4.381  &  4.413  & \\
      &  $1\,^1\Sigma^-$(2p,4p)     &  4.354  &  4.592  &  4.718  &  4.660  &  4.691  &  4.591  &            &  4.401  &  4.638  &  4.746  &  &  4.692  &  4.587  & \\
      &  $1\,^3\Pi$(2p,4s)          &  0.542  &  0.445  &         &         &  0.314  &  0.347  &  0.506(13) &  0.574  &  0.481  &         &  &  0.332  &  0.358  & \\
      &  $1\,^3\Sigma^+$($\sig$,4s) &  1.884  &  1.731  &         &         &  1.590  &  1.578  &  1.793(16) &  1.880  &  1.729  &         &  &  1.570  &  1.554  & \\
  \hline
  ZnS &  $1\,^1\Pi$(3p,4s)          &  0.769  &  0.755  &  0.778  &  0.774  &  0.735  &  0.701  &  0.816(3)  &  0.802  &  0.787  &  0.801  &  &  0.774  &  0.724  &  0.814(14)  \\
      &  $2\,^1\Sigma^+$($\sig$,4s) &  3.626  &  3.616  &  3.630  &  3.622  &  3.974  &  3.911  &  3.673(51) &  3.651  &  3.642  &  3.649  &  &  3.978  &  3.915  &  3.867(116) \\
      &  $1\,^1\Delta$(3p,4p)       &  4.181  &  4.162  &  4.213  &  4.200  &  4.198  &  4.191  &            &  4.225  &  4.204  &  4.242  &  &  4.231  &  4.216  &             \\
      &  $1\,^1\Sigma^-$(3p,4p)     &  4.225  &  4.213  &  4.279  &  4.267  &  4.315  &  4.252  &            &  4.252  &  4.238  &  4.294  &  &  4.335  &  4.258  &             \\
      &  $1\,^3\Pi$(3p,4s)          &  0.520  &  0.503  &         &         &  0.528  &  0.496  &            &  0.565  &  0.546  &         &  &  0.573  &  0.524  &             \\
      &  $1\,^3\Sigma^+$($\sig$,4s) &  2.343  &  2.302  &         &         &  2.311  &  2.304  &            &  2.351  &  2.306  &         &  &  2.314  &  2.306  &             \\
\end{tabular}
\end{ruledtabular}
\end{table*}
\end{squeezetable}

\begin{squeezetable}
\begin{table*}
\caption{Vertical excitation energies (in \si{\eV}) of the lowest doublet excited states of \ce{ScO}, \ce{ScS}, \ce{TiN}, and \ce{ZnH} at various levels of theory. 3, T, 4, Q, CAS, and NEV stand for CC3, CCSDT, CC4, CCSDTQ, CASPT2(IPEA), and PC-NEVPT2, respectively.}
\label{tab:doublets}
\begin{ruledtabular}
\begin{tabular}{llcccccccccccc}
  &            &  \mc{6}{c}{aug-cc-pVDZ}    &  \mc{6}{c}{aug-cc-pVTZ} \\
                        \cline{3-8}                 \cline{9-14}
  Mol.&  State        &  3    &  T  &  Q  &  CAS  &  NEV  &  FCI      &  3    &  T  &  Q  &  CAS  &  NEV  &  FCI \\
  \hline
  ScO &  $1\,^2\Pi$(4s,3d)      &  2.000  &  2.029  &  2.029  &  2.032  &  2.036  &  2.032(1)  &  1.998  &  2.028  &  2.028  &  2.037  &  2.040  &  2.037(3)  \\
      &  $1\,^2\Delta$(4s,3d)   &  2.248  &  2.023  &  2.084  &  1.824  &  1.829  &  2.107(1)  &  2.270  &  2.052  &  2.109  &  1.794  &  1.797  &  2.133(3)  \\
      &  $2\,^2\Sigma^+$(4s,3d) &  2.484  &  2.584  &  2.564  &  2.575  &  2.594  &  2.563(6)  &  2.482  &  2.584  &  2.564  &  2.569  &  2.591  &  2.572(19) \\
      &  $2\,^2\Pi$(2p,4s)      &  3.502  &  3.467  &  3.550  &  3.590  &  3.489  &            &  3.534  &  3.506  &  3.570  &  3.632  &  3.502  &            \\
  \hline
  ScS &  $1\,^2\Delta$(4s,3d)   &  1.376  &  1.185  &  1.262  &  0.981  &  0.974  &  1.300(2)  &  1.405  &  1.231  &  1.299  &  0.936  &  0.934  &  1.340(8) \\
      &  $1\,^2\Pi$(4s,3d)      &  1.579  &  1.450  &  1.485  &  1.495  &  1.466  &  1.495(8)  &  1.574  &  1.459  &  1.491  &  1.476  &  1.434  &  1.512(4) \\
      &  $2\,^2\Sigma^+$(4s,3d) &  1.663  &  1.572  &  1.591  &  1.637  &  1.590  &  1.590(1)  &  1.653  &  1.571  &  1.589  &  1.634  &  1.574  &  1.593(5) \\
      &  $2\,^2\Pi$(3p,4s)      &  2.161  &  2.021  &  2.070  &  2.134  &  2.140  &            &  2.185  &  2.063  &  2.093  &  2.215  &  2.338  &           \\
  \hline
  TiN &  $1\,^2\Delta$(4s,3d)     &  1.330  &  0.869  &  0.970  &  0.884  &  0.916  &  1.027(1)  &  1.386  &  0.916  &      &  0.840  &  0.874  &  1.066(5)  \\
      &  $2\,^2\Delta$($\sig$,3d) &  1.796  &  1.965  &  1.969  &  2.241  &  2.242  &  2.008(6)  &  1.863  &  2.000  &      &  2.227  &  2.222  &            \\
      &  $1\,^2\Pi$(4s,3d)        &  1.979  &  2.023  &  2.009  &  2.045  &  2.030  &            &  1.976  &  2.008  &      &  2.027  &  2.014  &            \\
  \hline
  ZnH &  $1\,^2\Pi$(4s,4p)          &  2.862  &  2.834  &  2.839  &  2.843  &  &  2.851(4)  &  2.874  &  2.846  &  &  2.862  &  &  2.882(17) \\
      &  $2\,^2\Sigma^+$($\sig$,4s) &  4.474  &  4.452  &  4.448  &  4.483  &  &  4.456(13) &  4.499  &  4.478  &  &  4.510  &  &  4.483(43) \\
      &  $3\,^2\Sigma^+$(4s,5s)     &  5.070  &  5.027  &  5.035  &  5.062  &  &  5.045(3)  &  5.097  &  5.056  &  &  5.097  &  &  5.115(5)  \\
      &  $4\,^2\Sigma^+$(4s,5p)     &  5.674  &  5.628  &  5.635  &  5.582  &  &  5.644(7)  &  5.691  &  5.647  &  &  5.598  &  &  5.689(21) \\
      &  $2\,^2\Pi$(4s,5p)          &  6.118  &  6.069  &  6.076  &  6.062  &  &  6.081(8)  &  6.147  &  6.100  &  &  6.106  &  &  6.126(12) \\
\end{tabular}
\end{ruledtabular}
\end{table*}
\end{squeezetable}

\begin{squeezetable}
\begin{table*}
\caption{TBEs (in \si{\eV}) in the aug-cc-pVDZ (AVDZ) and aug-cc-pVTZ (AVTZ) basis sets for the 11 diatomic molecules and the corresponding composite method to generate them.}
\label{tab:TBE}
\begin{ruledtabular}
\begin{tabular}{llcccccc}
      &               &  \mc{2}{c}{aug-cc-pVDZ}  &  \mc{2}{c}{aug-cc-pVTZ} & \mc{2}{c}{literature} \\
                            \cline{3-4}                   \cline{5-6}          \cline{7-8}
 Mol. &  State        &  TBE    &  Method        &  TBE    &  Method       & Exp.   & Th.   \\
  \hline
  ScH &  $1\,^1\Delta$(4s,3d)       &  0.606  &  CCSDTQ  &  0.672  &  CCSDTQ  & 0.519\fnm[\ScHa] & 0.509\fnm[\ScHb] \\
      &  $1\,^1\Pi$(4s,3d)          &  0.820  &  CCSDTQ  &  0.875  &  CCSDTQ  & 0.670\fnm[\ScHa] & 0.710\fnm[\ScHb] \\
      &  $2\,^1\Sigma^+$(4s,3d)     &  1.836  &  CCSDTQ  &  1.845  &  CCSDTQ  & 1.683\fnm[\ScHa] & 1.703\fnm[\ScHb] \\
      &  $2\,^1\Pi$(4s$^2$,3d$^2$)         &  2.181  &  CCSDTQ  &  2.215  &  CCSDTQ  & 2.089\fnm[\ScHa] & 2.151\fnm[\ScHb] \\
      &  $1\,^3\Delta$(4s,3d)       &  0.364  &  CCSDTQ  &  0.446  &  CCSDTQ  &                  & 0.225\fnm[\ScHb] \\
      &  $1\,^3\Pi$(4s,3d)          &  0.565  &  CCSDTQ  &  0.621  &  CCSDTQ  &                  & 0.430\fnm[\ScHb] \\
      &  $1\,^3\Sigma^+$(4s,3d)     &  0.820  &  CCSDTQ  &  0.851  &  CCSDTQ  &                  & 0.728\fnm[\ScHb] \\
      &  $2\,^3\Pi$(4s,3d)          &  1.832  &  CCSDTQ  &  1.817  &  CCSDTQ  &                  & 1.852\fnm[\ScHb] \\
  \hline
  ScO &  $1\,^2\Pi$(4s,3d)      &  2.032              &  FCI    &  2.037              &  FCI     & 2.057\fnm[\ScOa] & 2.070(6)\fnm[\ScOb] \\
      &  $1\,^2\Delta$(4s,3d)   &  2.107              &  FCI    &  2.133              &  FCI     & 1.915\fnm[\ScOa] & 1.950(5)\fnm[\ScOb] \\
      &  $2\,^2\Sigma^+$(4s,3d) &  2.563              &  FCI    &  2.572              &  FCI     & 2.559\fnm[\ScOc] &                     \\
      &  $2\,^2\Pi$(2p,4s)      &  3.550\fnm[\unsafe] &  CCSDTQ &  3.570\fnm[\unsafe] &  CCSDTQ  &                  &                     \\
  \hline
  ScF &  $1\,^1\Delta$(4s,3d)      &  0.797  &  FCI    &  0.903  &  FCI                                      & 0.568\fnm[\ScFa] & 0.478\fnm[\ScFd] \\
      &  $1\,^1\Pi$(4s,3d)         &  1.574  &  FCI    &  1.671  &  FCI/AVDZ + CCSDTQ/AVTZ $-$ CCSDTQ/AVDZ   & 1.184\fnm[\ScFa] & 1.107\fnm[\ScFd] \\
      &  $2\,^1\Sigma^+$(4s,3d)    &  2.346  &  CCSDTQ &  2.404  &  CCSDTQ                                   & 2.527\fnm[\ScFa] & 2.004\fnm[\ScFd] \\
      &  $2\,^1\Pi$(4s$^2$,3d$^2$) &  2.758  &  CCSDTQ &  2.837  &  CCSDTQ                                   & 2.750\fnm[\ScFa] & 2.521\fnm[\ScFd] \\
      &  $1\,^3\Delta$(4s,3d)      &  0.528  &  FCI    &  0.647  &  FCI                                      & 0.242\fnm[\ScFb] & 0.215\fnm[\ScFd] \\
      &  $1\,^3\Pi$(4s,3d)         &  1.037  &  FCI    &  1.109  &  FCI/AVDZ + CCSDT/AVTZ $-$ CCSDT/AVDZ     & 0.774\fnm[\ScFc] & 0.719\fnm[\ScFd] \\
      &  $1\,^3\Sigma^+$(4s,3d)    &  1.401  &  CCSDTQ &  1.459  &  CCSDTQ/AVDZ + CCSDT/AVTZ $-$ CCSDT/AVDZ  &                  & 1.073\fnm[\ScFd] \\
      &  $2\,^3\Pi$(4s$^2$,3d$^2$) &  2.469  &  CCSDTQ &  2.593  &  CCSDTQ/AVDZ + CCSDT/AVTZ $-$ CCSDT/AVDZ  &                  & 2.347\fnm[\ScFd] \\
  \hline
  ScS &  $1\,^2\Delta$(4s,3d)   &  1.300              &  FCI     &  1.340               &  FCI     &                  & 1.003\fnm[\ScSa] \\
      &  $1\,^2\Pi$(4s,3d)      &  1.495              &  FCI     &  1.512               &  FCI     & 1.375\fnm[\ScSb] & 1.418\fnm[\ScSa] \\
      &  $2\,^2\Sigma^+$(4s,3d) &  1.590              &  FCI     &  1.593               &  FCI     & 1.544\fnm[\ScSb] & 1.493\fnm[\ScSa] \\
      &  $2\,^2\Pi$(3p,4s)      &  2.070\fnm[\unsafe] &  CCSDTQ  &  2.093\fnm[\unsafe]  &  CCSDTQ  &                  &                  \\
  \hline
  TiN &  $1\,^2\Delta$(4s,3d)     &  1.027  &  FCI    &  1.066  &  FCI                                     & 0.934\fnm[\TiNa] & 0.946\fnm[\TiNb] \\
      &  $2\,^2\Delta$($\sig$,3d) &  2.008  &  FCI    &  2.043  &  FCI/AVDZ + CCSDT/AVTZ $-$ CCSDT/AVDZ    &                  &                  \\
      &  $1\,^2\Pi$(4s,3d)        &  2.009  &  CCSDTQ &  1.994  &  CCSDTQ/AVDZ + CCSDT/AVTZ $-$ CCSDT/AVDZ & 2.013\fnm[\TiNa] & 2.01\fnm[\TiNb]  \\
  \hline
  CuH &  $2\,^1\Sigma^+$(3d,4s)            &  3.051              &  FCI    &  3.080              &  FCI                                       & 2.905\fnm[\CuHa] & 3.009\fnm[\CuHb] \\
      &  $1\,^1\Delta$(3d,4s) &  3.718\fnm[\unsafe] &  CCSDTQ &  3.754\fnm[\unsafe] &  CCSDTQ                                    & 3.530\fnm[\CuHa] &                  \\
      &  $1\,^1\Pi$(3d,4s)         &  3.788              &  FCI    &  3.829              &  FCI                                       & 3.381\fnm[\CuHa] & 3.406\fnm[\CuHb] \\
      &  $2\,^1\Pi$(3d,4s)        &  5.470\fnm[\unsafe] &  CCSDTQ &  5.497\fnm[\unsafe] &  CCSDTQ                                    & 5.542\fnm[\CuHa] & 6.035\fnm[\CuHb] \\
      &  $3\,^1\Sigma^+$(3d,4p) &  5.713\fnm[\unsafe] &  CCSDTQ &  5.751\fnm[\unsafe] &  CCSDTQ                                    &                  & 3.349\fnm[\CuHb] \\
      &  $1\,^3\Sigma^+$(3d,4s)            &  2.514              &  FCI    &  2.550              &  FCI/AVDZ + CASPT2/AVTZ $-$ CASPT2/AVDZ    & 2.418\fnm[\CuHc] & 2.604\fnm[\CuHb] \\
      &  $1\,^3\Pi$(3d,4s)         &  3.522              &  FCI    &  3.582              &  FCI/AVDZ + CASPT2/AVTZ $-$ CASPT2/AVDZ    & 3.276\fnm[\CuHa] & 3.159\fnm[\CuHb] \\
      &  $1\,^3\Delta$(3d,4s) &  3.561\fnm[\unsafe] &  CCSDTQ &  3.618\fnm[\unsafe] &  CCSDTQ/AVDZ + CASPT2/AVTZ $-$ CASPT2/AVDZ & 3.492\fnm[\CuHa] &                  \\
      &  $2\,^3\Pi$(3d,4s)        &  4.762\fnm[\unsafe] &  CCSDTQ &  4.817\fnm[\unsafe] &  CCSDTQ/AVDZ + CASPT2/AVTZ $-$ CASPT2/AVDZ &                  & 5.209\fnm[\CuHb] \\
  \hline
  CuF &  $2\,^1\Sigma^+$(3d,4s)            &  2.561              &  FCI    &  2.572              &  FCI/AVDZ + CASPT2/AVTZ $-$ CASPT2/AVDZ    & 2.445\fnm[\CuFa] & 2.31\fnm[\CuClb] \\
      &  $1\,^1\Pi$(3d,4s)         &  2.751              &  FCI    &  2.793              &  FCI/AVDZ + CASPT2/AVTZ $-$ CASPT2/AVDZ    & 2.512\fnm[\CuFa] & 2.41\fnm[\CuClb] \\
      &  $1\,^1\Delta$(3d,4s) &  3.285\fnm[\unsafe] &  FCI    &  3.316\fnm[\unsafe] &  FCI/AVDZ + CASPT2/AVTZ $-$ CASPT2/AVDZ    &                  & 2.93\fnm[\CuClb] \\
      &  $2\,^1\Pi$(3d,4s)        &  5.914\fnm[\unsafe] &  CCSDTQ &  5.933\fnm[\unsafe] &  CCSDTQ/AVDZ + CASPT2/AVTZ $-$ CASPT2/AVDZ &                  &                  \\
      &  $1\,^3\Sigma^+$(3d,4s)            &  2.017              &  FCI    &  2.063              &  FCI/AVDZ + CASPT2/AVTZ $-$ CASPT2/AVDZ    & 1.808\fnm[\CuFa] & 1.81\fnm[\CuFc]  \\
      &  $1\,^3\Pi$(3d,4s)         &  2.421              &  FCI    &  2.471              &  FCI/AVDZ + CASPT2/AVTZ $-$ CASPT2/AVDZ    & 2.177\fnm[\CuFa] & 2.17\fnm[\CuFc]  \\
      &  $1\,^3\Delta$(3d,4s) &  2.937\fnm[\unsafe] &  CCSDTQ &  2.972\fnm[\unsafe] &  CCSDTQ/AVDZ + CASPT2/AVTZ $-$ CASPT2/AVDZ & 2.827\fnm[\CuFa] & 2.65\fnm[\CuFc]  \\
      &  $2\,^3\Pi$(3d,4s)        &  5.766\fnm[\unsafe] &  CCSDTQ &  5.800\fnm[\unsafe] &  CCSDTQ/AVDZ + CASPT2/AVTZ $-$ CASPT2/AVDZ &                  &                  \\
  \hline
 CuCl &  $2\,^1\Sigma^+$($\sig$,4s)  &  3.004\fnm[\unsafe]  &  CCSDTQ  &  3.065\fnm[\unsafe]  & CCSDTQ/AVDZ + CASPT2/AVTZ $-$ CASPT2/AVDZ & 2.848\fnm[\CuCla]  & 2.75\fnm[\CuClb] \\
      &  $1\,^1\Pi$(3d,4s)   &  3.008\fnm[\unsafe]  &  CCSDTQ  &  3.092\fnm[\unsafe]  & CCSDTQ/AVDZ + CASPT2/AVTZ $-$ CASPT2/AVDZ & 2.861\fnm[\CuCla]  & 2.78\fnm[\CuClb] \\
      &  $1\,^1\Delta$(3d,4s)    &  3.540\fnm[\unsafe]  &  CCSDTQ  &  3.479\fnm[\unsafe]  & CCSDTQ/AVDZ + CASPT2/AVTZ $-$ CASPT2/AVDZ &                    & 3.20\fnm[\CuClb] \\
      &  $1\,^3\Sigma^+$($\sig$,4s)  &  2.626\fnm[\unsafe]  &  CASPT2  &  2.692\fnm[\unsafe]  & CASPT2                                    & 2.352\fnm[\CuCla]  & 2.43\fnm[\CuFc]  \\
      &  $1\,^3\Pi$(3d,4s)   &  2.661\fnm[\unsafe]  &  CASPT2  &  2.845\fnm[\unsafe]  & CASPT2                                    & 2.540\fnm[\CuCla]  & 2.62\fnm[\CuFc]  \\
      &  $1\,^3\Delta$(3d,4s)    &  3.281\fnm[\unsafe]  &  CASPT2  &  3.229\fnm[\unsafe]  & CASPT2                                    & 3.134\fnm[\CuCla]  & 3.00\fnm[\CuFc]  \\
  \hline
  ZnH &  $1\,^2\Pi$(4s,4p)          &  2.851  &  FCI  &  2.882  &  FCI                                   & 2.90\fnm[\ZnHa] &  2.93\fnm[\ZnHb] \\
      &  $2\,^2\Sigma^+$($\sig$,4s) &  4.456  &  FCI  &  4.482  &  FCI/AVDZ + CCSDT/AVTZ $-$ CCSDT/AVDZ  & 3.42\fnm[\ZnHa] &  4.54\fnm[\ZnHb] \\
      &  $3\,^2\Sigma^+$(4s,5s)     &  5.045  &  FCI  &  5.115  &  FCI                                   & 5.09\fnm[\ZnHa] &  5.04\fnm[\ZnHb] \\
      &  $4\,^2\Sigma^+$(4s,5p)     &  5.644  &  FCI  &  5.663  &  FCI/AVDZ + CCSDT/AVTZ $-$ CCSDT/AVDZ  &                 &  5.70\fnm[\ZnHb] \\
      &  $2\,^2\Pi$(4s,5p)          &  6.081  &  FCI  &  6.126  &  FCI                                   &                 &  6.09\fnm[\ZnHb] \\
  \hline
  ZnO &  $1\,^1\Pi$(2p,4s)          &  0.771              &  FCI    &  0.791              &  FCI/AVDZ + CC4/AVTZ $-$ CC4/AVDZ      & 0.615\fnm[\ZnOa] & 0.54\fnm[\ZnOb] \\
      &  $2\,^1\Sigma^+$($\sig$,4s) &  3.417              &  FCI    &  3.423              &  FCI/AVDZ + CC4/AVTZ $-$ CC4/AVDZ      &                  & 3.75\fnm[\ZnOb] \\
      &  $1\,^1\Delta$(2p,4p)       &  4.611\fnm[\unsafe] &  CCSDTQ &  4.645\fnm[\unsafe] &  CCSDTQ/AVDZ + CC4/AVTZ $-$ CC4/AVDZ   &                  & 4.90\fnm[\ZnOb] \\
      &  $1\,^1\Sigma^-$(2p,4p)     &  4.660\fnm[\unsafe] &  CCSDTQ &  4.688\fnm[\unsafe] &  CCSDTQ/AVDZ + CC4/AVTZ $-$ CC4/AVDZ   &                  & 4.76\fnm[\ZnOb] \\
      &  $1\,^3\Pi$(2p,4s)          &  0.506              &  FCI    &  0.542              &  FCI/AVDZ + CCSDT/AVTZ $-$ CCSDT/AVDZ  & 0.305\fnm[\ZnOa] & 0.29\fnm[\ZnOb] \\
      &  $1\,^3\Sigma^+$($\sig$,4s) &  1.793              &  FCI    &  1.791              &  FCI/AVDZ + CCSDT/AVTZ $-$ CCSDT/AVDZ  & 1.875\fnm[\ZnOc] & 2.03\fnm[\ZnOb] \\
  \hline
  ZnS &  $1\,^1\Pi$(3p,4s)          &  0.816              &  FCI     &  0.814              &  FCI                                  &  & 0.682\fnm[\ZnSa] \\
      &  $2\,^1\Sigma^+$($\sig$,4s) &  3.622\fnm[\unsafe] &  CCSDTQ  &  3.641\fnm[\unsafe] &  CCSDTQ/AVDZ + CC4/AVTZ $-$ CC4/AVDZ  &  & 3.718\fnm[\ZnSa] \\
      &  $1\,^1\Delta$(3p,4p)       &  4.200              &  CCSDTQ  &  4.229              &  CCSDTQ/AVDZ + CC4/AVTZ $-$ CC4/AVDZ  &  & 4.279\fnm[\ZnSa] \\
      &  $1\,^1\Sigma^-$(3p,4p)     &  4.267              &  CCSDTQ  &  4.282              &  CCSDTQ/AVDZ + CC4/AVTZ $-$ CC4/AVDZ  &  & 4.239\fnm[\ZnSa] \\
      &  $1\,^3\Pi$(3p,4s)          &  0.503\fnm[\unsafe] &  CCSDT   &  0.546\fnm[\unsafe] &  CCSDT                                &  & 0.456\fnm[\ZnSa] \\
      &  $1\,^3\Sigma^+$($\sig$,4s) &  2.302\fnm[\unsafe] &  CCSDT   &  2.306\fnm[\unsafe] &  CCSDT                                &  & 2.266\fnm[\ZnSa] \\
\end{tabular}
\end{ruledtabular}
$^a${Unsafe TBE which means error possibly greater than \SI{0.043}{\eV}.}
$^b${0-0 energy from emission spectroscopy of Ref.~\onlinecite{Ram_1997}.}
$^c${$T_e$ energy from MRCI calculations of Ref.~\onlinecite{Tabet_2021}.}
$^d${Vertical energy obtained from the $T_e$ chemiluminescence spectroscopy of Refs.~\onlinecite{Chalek_1976,Chalek_1977} corrected by the vibrational term, as explained in Ref.~\onlinecite{Jiang_2021}, and averaged over the two spin-orbit components.}
$^e${Vertical energy (and statistical uncertainty) from FCIQMC calculations of Ref.~\onlinecite{Jiang_2021}.}
$^f${$T_e$ energy from Ref.~\onlinecite{Huber_1979}.}
$^g${$T_e$ energy from emission spectroscopy of Ref.~\onlinecite{Lebeaultdorget_1994}.}
$^h${$T_e$ energy from emission spectroscopy of Ref.~\onlinecite{Shenyavskaya_1993}.}
$^i${0-0 energy from emission spectroscopy of Ref.~\onlinecite{Shenyavskaya_1993}.}
$^j${$T_e$ energy from MRCI+Q calculations of Ref.~\onlinecite{Langhoff_1988}.}
$^k${$T_e$ energy from MRCI+Q calculations of Ref.~\onlinecite{Romeu_2018}.}
$^l${0-0 energy from emission spectroscopy of Ref.~\onlinecite{Gengler_2006}.}
$^m${$T_e$ energy from emission spectroscopy of Ref.~\onlinecite{Brabaharan_1985}.}
$^n${$T_e$ energy from MRCI calculations of Ref.~\onlinecite{Harrison_1996}.}
$^o${$T_e$ experimental energy from Ref.~\onlinecite{Huber_1979}.}
$^p${$T_e$ energy from MRCI+Q+DKH calculations of Ref.~\onlinecite{Xu_2019}.}
$^q${0-0 energy from photoelectron spectroscopy of Ref.~\onlinecite{Calvi_2007}.}
$^r${$T_e$ energy from absorption spectroscopy of Ref.~\onlinecite{Ahmed_1982}.}
$^s${$T_e$ energy from EOM-CCSD calculations of Ref.~\onlinecite{Guichemerre_2002}.}
$^t${$T_e$ energy from CCSD(T) calculations of Ref.~\onlinecite{Guichemerre_2002}.}
$^u${0-0 energy from fluorescence spectroscopy of Ref.~\onlinecite{Delaval_1987}.}
$^v${0-0 experimental energy from Ref.~\onlinecite{Huber_1979}.}
$^w${Vertical energy from MS-CASPT2 calculations of Ref.~\onlinecite{Suo_2017}.}
$^x${0-0 energy from photoelectron spectroscopy of Ref.~\onlinecite{Kim_2001}.}
$^y${0-0 energy from MRCI+Q calculations of Ref.~\onlinecite{Sakellaris_2010}.}
$^z${0-0 energy from photoelectron spectroscopy of Ref.~\onlinecite{Moravec_2001}.}
$^{aa}${0-0 energy from MRCI+Q calculations of Ref.~\onlinecite{Papakondylis_2011}.}
\end{table*}
\end{squeezetable}


\subsection{ScH and ScF}
\label{sec:Sc1}

Out of the eleven transition metal diatomics considered here, the excitation energies of \ce{ScH} present the fastest convergence along the CC series.
Already at the CC3 level, most energies lie within the desired chemical accuracy window ($\pm$\SI{1}{kcal/mol} or \SI{0.043}{\eV}), as shown in Fig.~ \ref{fig:conv}.
The largest difference (\SI{0.12}{\eV}) appears for the fourth singlet state, {$2\,^1\Pi$(4s$^2$,3d$^2$)}, which is acceptable since this state has a significant doubly-excited character, making CC3 less efficient.
The accuracy is significantly improved at the CCSDT level and beyond.
These outcomes are unsurprising, as only 4 electrons are correlated in our (large) frozen core approximation for \ce{ScH}.
Hence, CCSDTQ is equivalent to FCI.
In CASPT2 and NEVPT2, all active electrons are correlated, though within a subset of orbitals, and the computed excitation energies deviate more from the TBEs than CC3.
There is an average increase of \SI{0.04}{\eV} in the TBEs of \ce{ScH} when going from aug-cc-pVDZ to aug-cc-pVTZ.
It is worth mentioning that the $1\,^3\Delta$(4s,3d) state of \ce{ScH} has the lowest TBE of the \textsc{quest} database,
of only \SI{0.364}{\eV} for the aug-cc-pVDZ basis set and \SI{0.446}{\eV} for the aug-cc-pVTZ basis set.

Our vertical TBEs are compatible with the 0-0 experimental energies, \cite{Ram_1997} the former being higher by \SIrange{0.13}{0.21}{\eV}.
The TBEs are also close to the T$_e$ energies calculated at the multireference configuration interaction (MRCI) level, \cite{Tabet_2021}
appearing higher in energy by \SIrange{0.06}{0.22}{\eV}.
The only exception concerns the $2\,^3\Pi${(4s,3d)} state, whose TBE is lying lower by \SI{-0.03}{\eV},
possibly related to the occurrence of an avoided crossing with a higher-lying $3\,^3\Pi$ state. \cite{Tabet_2021}

Moving to \ce{ScF}, we first note that both $2\,^1\Pi$(4s$^2$,3d$^2$) and $2\,^3\Pi$(4s$^2$,3d$^2$) excited states are doubly-excited with respect to the ground state.
Although the number of active electrons jumps from 4 to 10, the convergence along the CC series remains quite fast, though not on par with the \ce{ScH} case.
CC3 nevertheless delivers chemically accurate excitation energies for most states, except for the $2\,^3\Pi$(4s$^2$,3d$^2$) state, whose energy is more overestimated than the others. 
In fact, third-order methods like CC3 and CCSDT produce fairly accurate excitation energies for the two doubly excited states, with the largest difference to the TBEs produced with CC3 and the aug-cc-pVDZ basis set for the triplet state (\SI{0.156}{\eV}).
This is somewhat surprising given the typically poorer performance of these methods in describing such excited states. \cite{Loos_2019c}
Ramping up to CCSDTQ produces excitation energies too low by only \SI{0.01}{\eV} for the states whose TBEs are obtained with CIPSI.
Considering the complete set of 11 transition metals investigated here, 
we find the TBEs to be somewhat larger with the aug-cc-pVTZ basis set, by \SI{0.04}{\eV} in average and up to \SI{0.08}{\eV}.
Taking into account the safe TBEs only, \ce{ScF} shows the most pronounced basis set effects from our set, with the largest increase in the TBEs of \SI{0.124}{\eV} for the $2\,^3\Pi$(4s$^2$,3d$^2$) state,
followed by \SI{0.119}{\eV} for the $1\,^3\Delta$(4s,3d) state.

The TBEs of \ce{ScF} can be correlated with the experimental $T_e$ and 0-0 energies \cite{Lebeaultdorget_1994,Shenyavskaya_1993}
as well as with the $T_e$ values calculated with MRCI plus Davidson correction (+Q). \cite{Langhoff_1988}
However, the overall largest discrepancies from the current set of transition metal diatomics are seen for this system.
Compared to experiment, the TBEs can be lower by \SI{0.12}{\eV} [$2\,^1\Sigma^+$(4s,3d)] or higher by \SI{0.49}{\eV} [$1\,^1\Pi$(4s,3d)].
This hints at a more significant vibrational relaxation in the excited states than in \ce{ScH}.

\subsection{CuH, CuF, and CuCl}
\label{sec:Cu1}

\ce{CuH} has 12 active electrons, whereas the halogen-containing compounds \ce{CuF} and \ce{CuCl} have 18 in the present (large) frozen core calculations.
For some excited states of \ce{CuH} and \ce{CuF}, safe TBEs of FCI quality could be attained.
For others (and for \ce{CuCl}), we rely on CCSDTQ or CASPT2 as the TBEs, which are therefore considered unsafe.
These systems present strong oscillations in the CC series, having the overall slowest convergence from our set of transition metal compounds.
In many cases, detailed below, even CCSDTQ is unable to produce values within the chemically accurate TBEs that stem from FCI calculations.
The unfavorable convergence profile is independent of the atom bonded to the \ce{Cu}, which is the culprit for the observed behavior.
The 4s$^1$ unpaired electron is strongly coupled to the 3d$^{10}$ shell, making correlation effects very pronounced in \ce{Cu} containing systems.
This contrasts to the \ce{Zn} atom, where the additional electron fills the 4s$^2$ shell, such that a mean-field approximation is a much more suitable starting point than for strongly correlated \ce{Cu} atom.

For some states of \ce{CuH}, the CIPSI calculations yield a small error bar and therefore directly provide trustworthy TBEs.
Accounting for these states only, the average absolute errors are smaller with CCSDTQ (\SI{0.05}{\eV}) than with CASPT2 (\SI{0.12}{\eV}), besides being more systematic with the former method.
We thus expect CCSDTQ to perform similarly better than CASPT2 for the other excited states, where FCI is unattainable, and for this reason, CCSDTQ is the method of choice for obtaining their TBEs.
Given the systematic underestimation of CCSDTQ with respect to the available FCI estimates, by \SIrange{0.02}{0.08}{\eV} with the aug-cc-pVDZ basis set,
the true TBEs are probably greater (by a comparable amount) than those obtained with CCSDTQ.
The same reasoning holds for \ce{CuF}.
The average error for the 5 states where FCI/aug-cc-pVDZ estimates are accessible is smaller for CCSDTQ (\SI{0.06}{\eV}) than for CASPT2 (\SI{0.12}{\eV}),
 the former method underestimating the TBEs by \SIrange{0.02}{0.09}{\eV}.
The CCSDTQ results for the remaining 3 excited states thus provide our TBEs, which are in turn expected to be a bit too low.
When only CCSDT results are available (which is the case of the triplet states of \ce{CuCl}), we rely on CASPT2 for the TBEs, based on the same argument.
Despite the missing FCI estimates for \ce{CuCl}, the similarity between its excited states and those of \ce{CuH} and \ce{CuF} makes us believe that CCSDTQ would also be more accurate than CASPT2 for this system.
However, we note that the CC error formally increases more rapidly with the number of electrons than the CASPT2 one.
Enlarging the basis set from aug-cc-pVDZ to aug-cc-pVTZ affects all TBEs of \ce{CuH} quite similarly, which increase by \SIrange{0.03}{0.06}{\eV}, averaging at \SI{0.04}{\eV}.
For \ce{CuF}, they increase by \SIrange{0.01}{0.05}{\eV}, with an average of \SI{0.03}{\eV}.

For the three \ce{Cu} containing compounds, the profile of the CC convergence is similar for most states.
CC3 provides fairly decent excitation energies (typically within \SI{0.2}{\eV} of the TBEs), considering its relatively low computational cost.
CCSDT often becomes less accurate and underestimates the TBEs, which are then overestimated by CC4.
Overall, CC3 is more accurate (and cheaper) than the higher-order CCSDT and CC4 models.
CCSDTQ is probably enough to achieve an accuracy of \SI{0.1}{\eV}.
To ensure chemical accuracy (\SI{0.043}{\eV}), at least pentuple excitations should likely be accounted for in CC models.

The three species present the overall very large discrepancies between CC4 and CCSDTQ and between CC3 and CCSDT, from \SIrange{0.2}{0.3}{\eV}.
This is considerably more than usually observed for typical excited states of organic species. \cite{Loos_2018a,Loos_2020f,Loos_2021c,Loos_2022b}
They also showed the overall largest differences between the TBEs and the various CC models.
For instance, the $1\,^1\Pi$(3d,4s) state of \ce{CuH} is one of the most challenging of our set.
With the aug-cc-pVDZ basis set, the TBE of \SI{3.788}{\eV} obtained with FCI is considered safe.
This TBE presents the largest difference to CCSDT (\SI{3.537}{\eV}), of \SI{-0.251}{\eV}, and to CC4 (\SI{4.072}{\eV}), of \SI{+0.284}{\eV},
and the second largest difference to CCSDTQ (\SI{3.707}{\eV}), of \SI{-0.081}{\eV}.
For the aug-cc-pVTZ basis set, the same state also presents the largest energy difference obtained with CCSDTQ (\SI{3.746}{\eV}) and the safe TBE from FCI (\SI{3.829}{\eV}), of \SI{-0.083}{\eV}.


Despite these difficulties in the convergence of the CC series, there is overall good agreement between our TBEs and the previous results for \ce{CuH}, \ce{CuF}, and \ce{CuCl}.
Starting with \ce{CuH}, our TBEs appear \SIrange{-0.04}{0.45}{\eV} relative to the $T_e$ experimental energies \cite{Huber_1979} [0-0 energy in the case of the $1\,^3\Sigma^+$(3d,4s) state \cite{Calvi_2007}].
They are also consistent with MRCI+Q calculations (with relativistic effects),
except for the $3\,^1\Sigma^+$(3d,4p) state, much higher in energy in all our calculations (TBE of \SI{5.751}{\eV} with the larger basis set)
than in the previous one (\SI{3.349}{\eV}). \cite{Xu_2019}
This state has also been assigned as $3\,^1\Sigma^+$ in Ref.~\onlinecite{Xu_2019}, so the reason for this discrepancy is unclear.

The 4 states of \ce{CuF} for which safe TBEs can be compared with experiment, \cite{Ahmed_1982} appear higher in energy than the experimentally obtained $T_e$ values from \SIrange{0.13}{0.29}{\eV}.
They are also consistent with previous CCSD(T) and EOM-CCSD calculations. \cite{Guichemerre_2002}
Our TBEs for the $1\,^3\Delta$(3d,4s) state are close to and in-between the experimental \cite{Ahmed_1982} and EOM-CCSD results. \cite{Guichemerre_2002}
For the $2\,^1\Pi$(3d,4s) and $2\,^3\Pi$(3d,4s) excited states, there is, to our knowledge, no previous data to compare our results with.

Our TBEs for \ce{CuCl} are also consistent with the 0-0 experimental values, \cite{Delaval_1987}
which are overestimated by \SIrange{0.09}{0.34}{eV}.
There is also a correspondence between our TBEs and the $T_e$ values obtained from CCSD(T) calculations for the 3 triplet states \cite{Guichemerre_2002}
and from EOM-CCSD calculations for the 3 singlet states, \cite{Guichemerre_2002} though with quite larger differences in general, from \SIrange{0.23}{0.40}{\eV}.

\subsection{ZnO and ZnS}
\label{sec:Zn1}

\ce{ZnO} and \ce{ZnS} have 18 active electrons in our large frozen core approximation, the largest number in our set of transition metal diatomics (along with \ce{CuF} and \ce{CuCl}).
Despite that, the convergence along the CC series can be considered satisfactory, being far superior to the case of \ce{Cu} containing compounds.

The CC estimates for both the $1\,^1\Pi$(2p,4s) and $1\,^3\Pi$(2p,4s) excited states of \ce{ZnO} always fall within the chemically accurate region, even with CC3.
Meanwhile, at least CCSDT is needed to reach the same level of accuracy for the $2\,^1\Sigma^+$($\sig$,4s) state,
whereas the $1\,^1\Delta$(2p,4p) and $1\,^1\Sigma^-$(2p,4p) states are more challenging and would require at least CCSDTQ, which provides our (unsafe) TBEs.
For the latter two excited states, CC3 produces excitation energies largely underestimated, by ca. \SI{0.3}{\eV}.
CCSDT significantly reduces the errors, bringing the energies from \SIrange{0.05}{0.07}{\eV} below their TBEs.
However, CC4 does not lead to further improvement, as it overestimates the TBEs by around the same amount, \SI{0.06}{\eV}.
CCSDTQ is taken as the TBE, but it is not clear whether this level of theory is enough to achieve chemical accuracy for these two excited states.

Our TBEs for \ce{ZnO} are consistent with the previous reports on the 0-0 energies. \cite{Kim_2001,Moravec_2001,Sakellaris_2010}
For the lowest singlet [$1\,^3\Pi$(2p,4s)] and triplet [$1\,^3\Pi$(2p,4s)] excited states,
they exceed the experimental values \cite{Kim_2001,Moravec_2001} by \SI{0.18}{\eV} and \SI{0.24}{\eV}, respectively,
and MRCI+Q calculations \cite{Sakellaris_2010} by \SI{0.25}{\eV} (both states).
These differences can be explained by the potential energy curves reported in Ref.~\onlinecite{Sakellaris_2010},
which indicate considerably stretched equilibrium geometries for these two excited states (more so for the $1\,^3\Pi$(2p,4s) state).
Conversely, our TBE for the $1\,^3\Sigma^+$($\sig$,4s) state is smaller than the experimental and the MRCI+Q 0-0 energy, by \SI{0.08}{\eV} and \SI{0.24}{\eV}, respectively.
Here, ground- and excited-state bond distances are actually similar,
and the underestimated TBE probably indicates that the excited-state zero-point vibrational energy is greater than the ground-state one. \cite{Sakellaris_2010}
There is no experimental data for the 3 higher-lying states, though our TBEs are systematically lower (by \SIrange{0.07}{0.33}{\eV})
than the values obtained in the MRCI+Q calculations. \cite{Sakellaris_2010}

For \ce{ZnS}, CC3 is able to provide excitation energies with virtual chemical accuracy.
It is rather surprising that it performs significantly better than for \ce{ZnO}, considering that both systems share the same excited states character and number of active electrons.
The $2\,^1\Sigma^+$($\sig$,4s) state is very well described already at the CC3 level, and the energy computed at higher-order CC levels fluctuates around the TBE (corresponding to the CCSDTQ value for this state).
Despite the proximity between the CCSDTQ and FCI results, the residual statistical uncertainty of the latter prevents us from claiming this TBE to be safe.
For the two triplet excited states, $1\,^3\Pi$(3p,4s) and $1\,^3\Sigma^+$($\sig$,4s), CCSDT represents our TBEs, which are also considered unsafe.
CCSDT slightly but systematically decreases the excitation energies of all states,
which worsens the comparison with the TBEs for the $1\,^1\Pi$(3p,4s), $1\,^1\Delta$(3p,4p), and $1\,^1\Sigma^-$(3p,4p) states.
For the latter two, one needs CC4 to obtain a significant reduction in the errors.

The behavior of the first singlet excited state of \ce{ZnS}, $1\,^1\Pi$(3p,4s), stands out.
Along the CC series (CC3, CCSDT, CC4, and CCSDTQ), its aug-cc-pVDZ excitation energy oscillates between \SI{0.755}{\eV} and \SI{0.778}{\eV}, a narrow interval of \SI{0.023}{\eV}.
However, the apparent convergence can be ruled out with our FCI estimate of \SI{0.816}{\eV}, which represents our TBE.
This value lies \SI{0.042}{\eV} above the CCSDTQ result and is significantly greater than the \SI{0.003}{\eV} statistical uncertainty of the FCI result.
A similar trend is observed for the aug-cc-pVTZ basis set, though with a substantially smaller gap (\SI{0.013}{\eV}) between FCI and CC4 (CCSDTQ is beyond computational reach).
The convergence profile of the CC series is overall similar with both basis sets, somewhat less so than observed for \ce{ZnO}.
The TBEs of both \ce{ZnO} and \ce{ZnS} are slightly larger with the aug-cc-pVTZ basis set, by \SI{0.02}{\eV} in average.

Our computed TBEs for \ce{ZnS} are once again close to previous MRCI+Q results for the 0-0 energies, \cite{Papakondylis_2011} with differences between \SIrange{-0.08}{0.13}{\eV},
and the TBEs appear lower in energy for 2 out of the 6 excited states, $2\,^1\Sigma^+$($\sig$,4s) and $1\,^1\Delta$(3p,4p).
To the best of our knowledge, \ce{ZnS} is the single compound investigated here for which there is no experimental excited-state data.

\subsection{ScO and ScS}
\label{sec:Sc2}

Both \ce{ScO} and \ce{ScS} are radicals having 9 active electrons in our calculations.
For most states, the CC series quickly converges to the TBEs, though less than in the closed-shell \ce{ScH} and \ce{ScF} derivatives.
In particular, the CC series is overall faster for \ce{ScF} than for \ce{ScO}.
This reflects the impact of the additional electron from the \ce{F} atom (compared with the \ce{O} atom) closing the 3p shell..
The effect is small however, when compared to the case where the single electron difference is associated with the transition metals, as shown earlier for \ce{CuH} and \ce{ZnH}.

The $1\,^2\Delta$(4s,3d) excited states of both \ce{ScO} and \ce{ScS} are the most challenging for these systems.
CC3 and CCSDT display deviations of around \SI{0.1}{\eV} with respect to the TBE, by either overestimating or underestimating it, respectively.
The description of these excited states only becomes chemically accurate at the CCSDTQ level.
For both species, the CC convergence profile is quite insensitive to the basis set.

Our TBE for the $1\,^2\Pi$(4s,3d) state of \ce{ScO} (\SI{2.037}{\eV}) very closely matches both
the vertical excitation energy obtained with accurate FCI quantum Monte Carlo (FCIQMC) calculations (\SI{2.070}{\eV}), \cite{Jiang_2021}
and the estimated experimental vertical energy (\SI{2.057}{\eV}), based on the $T_e$ of Refs.~\onlinecite{Chalek_1976,Chalek_1977}
corrected for vibrational effect according to Ref.~\onlinecite{Jiang_2021}.
In turn, the differences are greater for the $1\,^2\Delta$(4s,3d) state, with our TBE of \SI{2.133}{\eV} appearing higher in energy than both the FCIQMC one (\SI{1.950}{\eV}) \cite{Jiang_2021}
and the estimated experimental vertical energy (\SI{1.915}{\eV}). \cite{Chalek_1976,Chalek_1977,Jiang_2021}
Our TBEs place the $1\,^2\Delta$(4s,3d) state higher in energy than the $1\,^2\Pi$(4s,3d) state, in contrast to the available experimental and theoretical results.
This is not too serious, however, as their energy gap is small, probably around \SI{0.1}{\eV}.
As for the $2\,^2\Sigma^+$(4s,3d) state, our TBE differs from the experimental $T_e$ value \cite{Huber_1979} by \SI{0.013}{\eV} only.
There is no previous data for the higher-lying $2\,^2\Pi$(2p,4s) excited state as far as we know.

The TBEs for the $1\,^2\Pi$(4s,3d) and $2\,^2\Sigma^+$(4s,3d) excited states of \ce{ScS} are consistent with both the available 0-0 experimental values \cite{Gengler_2006}
and with MRCI+Q $T_e$ calculations, \cite{Romeu_2018} being higher in energy by \SIrange{0.05}{0.14}{\eV}.
Whereas the lowest-lying excited state of \ce{ScS}, $1\,^2\Delta$(4s,3d) remains to be observed experimentally,
our TBE is higher than its MRCI+Q counterpart \cite{Romeu_2018} by \SI{0.34}{\eV}, a larger difference than for the two higher-lying excited states.
At least to some extent, this can be explained by the larger equilibrium bond distance of the first excited state, based on the MRCI+Q potential energy curves. \cite{Romeu_2018}

\subsection{TiN and ZnH}
\label{sec:TiZn2}

\ce{TiN} and \ce{ZnH} are radical species, having 9 and 13 active electrons in our calculations, respectively.
The first excited state of \ce{TiN}, $1\,^2\Delta$(4s,3d), shows a slow convergence along the CC series.
Compared to the TBE, its excitation energy is significantly overestimated at the CC3 level (\SIrange{0.30}{0.32}{\eV}, depending on the basis set), becoming underestimated at CCSDT level (\SIrange{0.15}{0.16}{\eV}).
The error further decreases with CCSDTQ, though not enough to reach chemical accuracy, as the computed energy appears \SI{0.06}{\eV} below the TBE with the aug-cc-pVDZ basis set.
For the next state, $2\,^2\Delta$($\sig$,3d), CC3 also starts off with a large error, this time undeshooting the TBE by \SIrange{0.18}{0.21}{\eV}.
With the aug-cc-pVDZ basis set, the excitation energy obtained with CCSDT (\SI{1.965}{\eV}) and CCSDTQ (\SI{1.969}{\eV}) varies by only \SI{0.004}{\eV}, suggesting fairly converged results.
However, the TBE of \SI{2.008}{\eV} (obtained with FCI) still lies \SI{0.039}{\eV} above the CCSDTQ energy.
For both $1\,^2\Delta$(4s,3d) and $2\,^2\Delta$($\sig$,3d) states of \ce{TiN}, excitations beyond quadruples seem needed to effectively reach chemical accuracy.
The higher-lying excited state, $1\,^2\Pi$(4s,3d), displays a very small error already at the CC3 level, becoming tinier at the higher CC levels.
We further notice that the convergence profile up to CCSDT is quite insensitive to the choice of basis sets.

The TBEs for \ce{TiN} are very close to the available $T_e$ energies obtained experimentally \cite{Brabaharan_1985} and with MRCI calculations. \cite{Harrison_1996}
Compared to the previously reported values, the TBE for the $1\,^2\Delta$(4s,3d) state is larger by \SIrange{0.12}{0.13}{\eV},
whereas for the $1\,^2\Pi$(4s,3d) state, it is smaller by \SI{0.07}{\eV}.
The $2\,^2\Delta$($\sig$,3d) excited state, although very close lying in energy to the $1\,^2\Pi$(4s,3d) state, has not been considered in the previous experimental and theoretical studies.

\ce{ZnH} presents one of the most favorable CC convergence profiles with the aug-cc-pVDZ basis set.
CC3 produces chemically accurate excitation energies with a slight overestimation with the double-$\zeta$ basis set.
CCSDT is very accurate with the aug-cc-pVDZ basis set (within \SI{0.02}{\eV} of the TBEs),
but slightly less with the aug-cc-pVTZ basis set (with deviations up to \SI{0.06}{\eV}).
That is a sizeable basis set effect not seen in the other transition metal diatomics.
CCSDTQ underestimates the TBEs by \SIrange{0.005}{0.015}{\eV} only.

There is very good agreement between the TBEs of \ce{ZnH} and the vertical excitation energies computed with multistate CASPT2 (MS-CASPT2) calculations, \cite{Suo_2017}
with average absolute deviations of \SI{0.06}{\eV} for the five excited states considered here.
Our CASPT2 results are even more accurate, with average absolute deviations of \SI{0.03}{\eV}.
The TBEs are also very close to the available experimental data \cite{Huber_1979} for the $1\,^2\Pi$(4s,4p) and $3\,^2\Sigma^+$(4s,5s) 0-0 energies,
respectively underestimated and overestimated by only \SI{0.02}{\eV}.
A much larger deviation of \SI{1.06}{\eV} is seen for the $2\,^2\Sigma^+$($\sig$,4s) state, probably reflecting its more stretched equilibrium bond distance. \cite{Huber_1979,Jamorski_1994,Suo_2017}

\subsection{Global statistics}
\label{sec:global}

\begin{figure}
  \includegraphics[width=\linewidth]{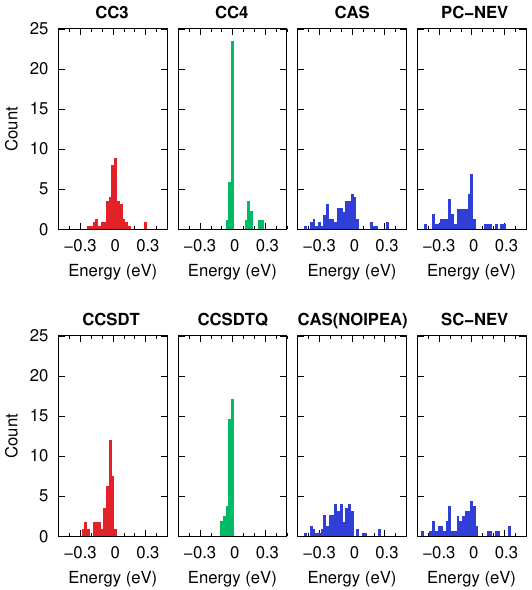}
  \caption{Distribution of the errors in excitation energies with respect to the safe TBEs of Table \ref{tab:TBE}, with corresponding statistical errors presented in Table \ref{tab:errors},
	where CAS and NEV stand for CASPT2 and NEVPT2, respectively.}
  \label{fig:dist}
\end{figure}

We computed the mean signed error (MSE), mean absolute error (MAE), and root-mean-square error (RMSE), gathered in Table \ref{tab:errors}, for both CC and multiconfigurational methods considered here.
They were evaluated with respect to the TBEs displayed in Table \ref{tab:TBE},
including results for both basis sets and excluding the unsafe excited states (having errors potentially greater than \SI{0.043}{\eV}).
Figure \ref{fig:dist} shows the corresponding distribution of errors in the excitation energies.
We collect the results from both basis sets because the main trends, discussed in the following, are rather insensitive to the choice of the basis set.
The individual results for each one can be found in the {\SupInf}.

\begin{table}[ht!]
\caption{Mean signed error (MSE), mean absolute error (MAE), and root-mean-square error (RMSE), in units of eV, with respect to the TBEs for all the states assigned as safe in Table \ref{tab:TBE}.}
\label{tab:errors}
\begin{ruledtabular}
\begin{tabular}{lddddd}
Method            & \mc{1}{c}{\#} & \mc{1}{c}{MSE} & \mc{1}{c}{MAE} & \mc{1}{c}{RMSE} \\
\hline
CC3               & 90  & +0.02 & 0.06 & 0.09 \\
CCSDT             & 90  & -0.05 & 0.06 & 0.09 \\
CC4               & 34  & +0.05 & 0.05 & 0.10 \\
CCSDTQ            & 63  & -0.02 & 0.02 & 0.03 \\
\hline
CASPT2 (IPEA)     & 90  & -0.08 & 0.12 & 0.16 \\
CASPT2 (no IPEA)  & 90  & -0.11 & 0.13 & 0.17 \\
PC-NEVPT2         & 64  & -0.08 & 0.12 & 0.16 \\
SC-NEVPT2         & 64  & -0.07 & 0.14 & 0.18 \\
\end{tabular}
\end{ruledtabular}
\end{table}

CC3 displays a fairly normal distribution of errors, centered around zero (MSE of \SI{+0.02}{\eV}) and the associated MAE is \SI{0.06}{\eV} only.
By fully accounting for the triple excitations, CCSDT produces a negatively skewed distribution, with the MSE moving further away from zero (\SI{-0.05}{\eV}), and the MAE remaining at \SI{0.06}{\eV}.
Moving to CC4 only slightly reduces the MAE to \SI{0.05}{\eV}, whereas the underlying distribution becomes positively skewed with a MSE of \SI{+0.05}{\eV}.
CCSDTQ significantly reduces the errors, with a MAE of only \SI{0.02}{\eV} and a somewhat negatively skewed distribution with a MSE of \SI{-0.02}{\eV}.
These very small errors would be expected, given that one-third of our safe TBEs stem from CCSDTQ calculations.
Excluding these TBEs slightly shifts the MAE to \SI{0.03}{\eV} and the MSE to \SI{-0.03}{\eV}.

In turn, the four different multiconfigurational approaches globally yield less accurate excitation energies than the CC models.
The errors are quite comparable among them, with the MAEs lying between \SIrange{0.12}{0.14}{\eV}, MSEs of \SIrange{-0.11}{-0.08}{\eV}, and negatively skewed distribution of the errors for all methods.
CASPT2 with the IPEA shift is the most accurate out of the four, yet by a small margin.
The effect of the IPEA shift is rather small, decreasing the MAE by \SI{0.01}{\eV} and making the MSE less negative by \SI{0.03}{\eV}.
Similarly, the errors of PC-NEVPT2 and SC-NEVPT2 differ by no more than \SI{0.02}{\eV}.
The comparable statistics obtained with the multiconfigurational methods endorses the choice of active spaces.

The accuracy of the tested CC models is overall independent of the basis set and is comparable for singlets, doublets, and triplets.
They are also similar across the different spatial symmetries, except for CC3, which performs better for $\Sigma$ and $\Pi$ (MAE of \SI{0.05}{\eV}) than for $\Delta$ (MAE of \SI{0.09}{\eV}) excited states.
In contrast, the multiconfigurational methods display more pronounced differences, improving towards states of lower angular momentum.
CASPT2/PC-NEVPT2 show a MAE of \SI{0.21}{\eV}/\SI{0.20}{\eV} for $\Delta$ states, decreasing to \SI{0.12}{\eV}/\SI{0.10}{\eV} for $\Pi$ states, and down to \SI{0.08}{\eV}/\SI{0.09}{\eV} for $\Sigma$ states.
The multiconfigurational methods also deliver somewhat more accurate results for triplet than for singlet excited states,
with respective MAEs of \SI{0.12}{\eV} and \SI{0.15}{\eV} according to CASPT2, and of \SI{0.11}{\eV} and \SI{0.14}{\eV} based on PC-NEVPT2.
Finally, the results are closer to the TBEs with the smaller basis set.
Both CASPT2 and PC-NEVPT2 have a MAE of \SI{0.11}{\eV} with the aug-cc-pVDZ basis set, which increases to \SI{0.14}{\eV} with the larger aug-cc-pVTZ basis set.
The same trends concerning basis sets and spatial/spin symmetries are observed with or without the IPEA shift and also for both PC-NEVPT2 and SC-NEVPT2.

It is interesting to compare the statistical errors for the present transition metal compounds with those for organic molecules obtained in the previous sets of the \textsc{quest} database.
For the transition metals, CC3 offers a MAE of \SI{0.06}{\eV}.
Although acceptable for most purposes, such error is greater than those previously found for other types of excited states.
By comparing with previous \textsc{quest} subsets for which CCSDTQ or FCI TBEs are available,
we find that the MAEs of CC3 become progressively smaller for
the radicals of \textsc{quest}\#4 \cite{Loos_2020f} (\SIrange{0.05}{0.06}{\eV}),
the small molecules of \textsc{quest}\#1 \cite{Loos_2018a} (\SI{0.03}{\eV}),
and the exotic molecules of \textsc{quest}\#4 \cite{Loos_2020f} (\SI{0.01}{\eV}).
Even though CC3 performs very well in absolute terms, the MAEs for each type of transition span a range of \SIrange{0.01}{0.06}{\eV}, which is large in relative terms.
The accuracy is excellent for typical transitions of organic molecules, though less so for radicals
and for transitions whose states present pronounced multiconfigurational character, as the transition metal derivatives surveyed here.

A similar comparison of CASPT2 and NEVPT2 reveals a different picture.
For the transition metal diatomics, the MAE of \SIrange{0.12}{0.14}{\eV}
is virtually the same as obtained for the medium-sized organic molecules of \textsc{quest}\#3 \cite{Loos_2020c} with NEVPT2 (\SI{0.13}{\eV}).
Even though these results certainly depend on the choice of active space, they highlight the versatility of CASPT2 and NEVPT2 methods in handling excited states with varying multiconfigurational characters while providing very similar levels of accuracy.
This contrasts with the case of CC3, whose accuracy is more dependent on the type of transitions.
However, even for the challenging transitions in transition metal diatomics, the single-reference CC3 still outperforms the multiconfigurational alternatives.
Interestingly, the IPEA shift has apparently a smaller impact on the transition metal diatomics than on organic systems. \cite{Sarkar_2022,Boggio-Pasqua_2022}
In addition, the CASPT2 and NEVPT2 excitation energies are generally very similar for the systems considered here.
These observations further support our active spaces choices.

\section{Conclusion}
\label{sec:conclusion}

We have presented highly-accurate vertical excitation energies for 67 excited states of 11 transition metal diatomic molecules, comprising 4 different fourth-row elements (\ce{Sc}, \ce{Ti}, \ce{Cu}, and \ce{Zn}). To this end, we employed state-of-the-art excited-state methods, including selected CI, high-order equation-of-motion CC (CC3, CCSDT, CC4, and CCSDTQ), and multiconfigurational (CASPT2 and NEVPT2) methods. These calculations allowed us to provide non-relativistic theoretical best estimates (based on the aug-cc-pVDZ and aug-cc-pVTZ basis sets) for the excitation energies of 67 states, 45 of which should be chemically accurate (errors less than \SI{0.043}{\eV} or \SI{1}{kcal/mol}). These TBEs were compared with previous experimental and theoretical results. This contribution establishes the eighth subset of the \textsc{quest} database, the first comprising transition metals.

The convergence of the CC series toward the TBE shows a pronounced dependence on the system. It is quite favorable for the \ce{Sc}-containing species, \ce{ScH}, \ce{ScF}, \ce{ScO}, and \ce{ScS} (somewhat less so for the latter two, open-shell radicals), followed by the \ce{Zn}-containing species, \ce{ZnH}, \ce{ZnO}, and \ce{ZnS}, although \ce{ZnO} presents a few challenging excited states. \ce{TiN} presents a slower convergence profile, whereas the \ce{Cu}-containing compounds, \ce{CuH}, \ce{CuF}, and \ce{CuCl} proved to be the most challenging systems from the present set. This trend can be rationalized based on the occupancy of the 3d and 4s shells of the transition atom. Moving one position towards the center of the periodic table (from \ce{Sc} to \ce{Ti} and from \ce{Zn} to \ce{Cu}) increases the half-filled character of the shells, making the electronic correlation problem harder to tackle.

Despite the challenging multiconfigurational character of many excited states, CC3 performs surprisingly well, with a MAE of \SI{0.06}{\eV}, not significantly more than observed for transitions of small organic systems (\SIrange{0.01}{0.06}{\eV}). The higher-order CCSDT and CC4 levels produce comparable MAEs, although their corresponding error distributions are negatively and positively skewed, respectively. A further reduction in the errors only comes at the CCSDTQ level, with a MAE of \SI{0.02}{\eV}. In turn, the multiconfigurational methods are less accurate than CC3, with MAEs of \SIrange{0.12}{0.14}{\eV}. Yet, we found quite consistent results with both forms of NEVPT2 and CASPT2, and a small effect introduced by the IPEA shift. Overall, if an accuracy of around \SIrange{0.1}{0.2}{\eV} is acceptable, CC3 would be recommended. Otherwise, CCSDTQ is needed to achieve chemical accuracy, though it is still insufficient for the most difficult cases.

While the current reference vertical energies offer significant value, it is important to note that they do not take into account relativistic effects. \cite{Saue_2020} Incorporating such effects, which can significantly affect vertical transition energies in some cases, would be both a logical and formidable undertaking that holds substantial potential value for the electronic structure community.

\acknowledgements{
This work was performed using HPC resources from CALMIP (Toulouse) under allocation 2023-18005.
D.J.~is indebted to the CCIPL/GliCID computational center installed in Nantes for a generous allocation of computational time.
This project has received funding from the European Research Council (ERC) under the European Union's Horizon 2020 research and innovation programme (Grant agreement No.~863481).}

\section*{Supporting Information}
Additional statistical measures for each basis set and various types of excited states, comparison between small and large frozen-core calculations, additional CASSCF, CASPT2 (with and without IPEA shift), and (partially- and strongly-contracted) NEVPT2 data, as well as the detailed description and specification of the active space for each molecule and each transition.


\bibliography{TM,acs-TM}

\end{document}